\documentclass[fleqn,usenatbib]{mnras}

\usepackage{color}
\usepackage{newtxtext,newtxmath,comment}

\usepackage[T1]{fontenc}

\DeclareRobustCommand{\VAN}[3]{#2}
\let\VANthebibliography\thebibliography
\def\thebibliography{\DeclareRobustCommand{\VAN}[3]{##3}\VANthebibliography}

\usepackage{graphicx}	
\usepackage{amsmath}	

\title[HCF4]{HCF4 long}

\newcommand\cm{{\rm\thinspace cm}}

\newcommand\erg{{\rm\thinspace erg}}

\newcommand\K{{\rm\thinspace K}}
\newcommand\keV{{\rm\thinspace keV}}
\newcommand\km{{\rm\thinspace km}}

\newcommand\Msun{\hbox{$\rm\thinspace M_{\odot}$}}

\newcommand\pc{{\rm\thinspace pc}}

\newcommand\s{{\rm\thinspace s}}
\newcommand\yr{{\rm\thinspace yr}}

\newcommand\cmsq{\hbox{$\cm^2\,$}}

\newcommand\ppccu{\hbox{$\pc^{-3}\,$}}

\newcommand\ergps{\hbox{$\erg\s^{-1}\,$}}

\newcommand\kmps{\hbox{$\km\s^{-1}\,$}}

\newcommand\Msunpyr{\hbox{$\Msun\yr^{-1}\,$}}

\newcommand\pcmsq{\hbox{$\cm^{-2}\,$}}

\newcommand\psqcm{\hbox{$\cm^{-2}\,$}}

\title[Hidden Cooling Flows V]{Hidden (absorbed) Cooling Flows V: Groups and Galaxies including Spirals }

\author[A. C. Fabian et al.]{
A. C. Fabian,$^{1}$\thanks{E-mail: acf@ast.cam.ac.uk }, J.S. Sanders$^{2}$, G.J. Ferland$^{3}$, H.R. Russell$^{4}$, B.R. McNamara$^{5}$, C. Pinto$^{6}$ and S.A. Walker$^{7}$
\\
$^{1}$Institute of Astronomy, University of Cambridge, Madingley Road, Cambridge CB3 0HA, UK\\
$^{2} $Max-Planck-Institut fur extraterrestrische Physik, Giessenbachstrasse 1, 85748 Garching, Germany\\
$^{3} $Department of Physics, University of Kentucky, Lexington KY 40506, USA\\
$^{4} $School of Physics \& Astronomy, University of Nottingham, University Park, Nottingham NG7 2RD, UK\\
$^{5} $Department of Physics and Astronomy, University of Waterloo, 200 University Avenue West, Waterloo, ON N2L 3G1, Canada\\
$^{6}$INAF-IASF Palermo, Via U. La Malfa 153, I-90146 Palermo, Italy\\
$^{7}$Department of Physics and Astronomy, The University of Alabama in Huntsville, Huntsville, AL 35899, USA\\ 
}
\date{Accepted XXX. Received YYY; in original form ZZZ}
\pubyear{2024}
\begin{document}
\label{firstpage}
\pagerange{\pageref{firstpage}--\pageref{lastpage}}
\maketitle
\begin{abstract}
Cooling flows are observed in X-ray studies of the centres of cool core clusters, galaxy groups and individual elliptical galaxies. They are partly hidden from direct view by embedded cold gas so have been called Hidden Cooling Flows. X-ray spectra from the XMM RGS reveal emission from hot gas modified  by photoelectric absorption by cold gas intrinsic to the flow. Here we present the spectral analysis of 6 more low redshift galaxy groups ranging from the nearest fossil group to 2 groups hosting bright radio sources. All reveal absorbed cooling flows. AGN feedback is ineffective in heating the inner cooling gas in groups and elliptical galaxies. We have extended the analysis to include 3 nearby spiral galaxies (the Sombrero, Whirlpool and Sculptor galaxies). They have similar absorbed soft X-ray spectra to  elliptical galaxies and may also host cooling flows of 0.3 to $1.1\Msunpyr$ in their CircumGalactic Medium.

\end{abstract}



\section{Introduction}

Many galaxies reside in clusters and groups within which most of the baryons lie in a hot diffuse medium. The gas has been heated earlier to temperatures of 10 million K and above by gravitational infall in the deep gravitational wells provided by dominant dark matter. Many have cool cores in which the temperature drops inward as the gas cools due to the emission of bremsstrahlung X-rays. This situation indicates the presence of a cooling flow in which the gas continues to cool inward as the  density rises to support the weight of the overlying gas (see e.g. \citep{Fabian1994rev}). 

Initial observational analyses of soft X-ray spectra provided by the Reflection Grating Spectrometers (RGS) on the XMM X-ray Observatory however did not support the presence of steady cooling flows (\cite{Kaastra2001}, \cite{Peterson2001}, \cite{Peterson2003}, \cite{Peterson2006}, \cite{Liu2019}). Active Galactic Nucleus (AGN) Feedback of energy from an accretion central black hole was proposed to balance the energy loss due to radiative cooling (see \citep {McNamara2012, Fabian2012} for reviews). This balance appeared to be very tight since there is no, or little, evidence of normal star formation at the centre of many cluster and group brightest central galaxies (e.g.\citep   {McDonald2018, Fabian24, Tamhane25}). 

We have been questioning how such a tight balance can arise and persist. We have explored the possibility that cooling flows are taking place but hidden from  view by photoelectric absorption in cool gas, including gas which was earlier part of the cooling flow (HCFI, II, III, IV): \cite{Fabian22}, \cite{Fabian23}, \cite{Fabian23b}, \cite{Fabian24b} in 29 cool core clusters. In this model, the absorbed energy then emerges in the Far Infrared and the absorbing gas is dispersed throughout the hot component \citep{Allen1997}\footnote{A similar situation of spatially distributed emission and absorption is observed and modelled in the winds of massive O stars (e.g.\cite{Cassinelli79, Stewart81, Leutenegger10, Cohen21}.} 

Deep central temperature drops and short central cooling times are also common in groups of galaxies (see e.g \cite{Lakhchaura2018}). In previous work, We have examined XMM RGS spectra for Hidden Cooling Flows in 8 groups (NGC 5044, NGC 1550, NGC1600, NGC3091 (Hickson 42), NGC5813, NGC5846, NGC533 and A3581) and here we look at 6 more (NGC 6482, IC4296, NGC4325, NGC3411, NGC 4261 and NGC2300. They have been selected to show clear FeXVII emission at 15 and 17\AA\ in RGS quick-look data (XMM X-ray Science Archive). We have also studied the RGS spectra of 4 elliptical galaxies (M84, M49, NGC720 nd Mrk 1216) from the Virgo cluster and the field in the above HCFII and III papers as well as 7 nearby elliptical galaxies in \citep{Ivey24}. This makes a total of 50 objects. 

We have now extended the work to include 3 nearby spiral galaxies with prominent FeXVII line emission: the Sombrero Galaxy M104, the Whirlpool Galaxy M51 and the Sculptor Galaxy NGC253. Unlike groups, clusters and even massive elliptical galaxies. Unlike groups of galaxies, the spirals have no outer reservoir of 1 keV gas to fuel a standard cooling flow and their halo gas is at a temperature of less than 0.7 keV.  What we are testing is whether that gas contains a component {\it consistent} with a cooling flow. Several theoretical studies have recently proposed that cooling flows could be operating in spiral galaxies (e.g. \cite{Stern19}, \cite{Dutta22}, \cite{Sultan25}).

\section{The XMM RGS analysis} 

The XMM RGS \citep{denHerder2001} provides the spectrum from an effective slit of typically 90, 95 and 99 percent in the cross-dispersion direction, corresponding to 0.8, 1.7 and 3.4 arcmin width. Here we use the 95 percent aperture data. RGS 1 and 2 have dead chips covering 10.4--13.8 \AA\ and 20--24 \AA\ respectively. We concentrate on the 1--20\AA\ band where the RGS background is lowest. The spectral analysis is carried out in \textsc{XSPEC} \citep{Arnaud1996}.

The total spectral model consists of Galactic absorption applied overall to a single temperature \textsc{APEC} component plus a multilayer absorption model \textsc{ mlayerz} $=$\ ($(1-\exp(-\sigma N_{\rm H}))/(\sigma N_{\rm H}))$, where $\sigma$ is the photoelectric cross section at the relevant energy and $N_{\rm H}$ is the total column density) applied to a cooling flow model \textsc{mkcflow}. ( In \textsc{XSPEC }, the model function \textsc{mlayerz} is defined with the expression \textsc{mdefine}.) The higher temperature of the cooling flow is fixed to equal that of the single temperature outer component and the lower temperature is $0.1\keV$\footnote{It is assumed that the gas continues to cool below $0.1\keV$ in the host galaxy, but does not further emit any soft X-rays detectable in the RGS.}. As the spectrum is from an extended object, gaussian smoothing is applied separately to the 2 components to account for the spatial-spectral blurring which occurs. In earlier work we have allowed for the possibility of partial covering of the X-ray emitting gas. As most of those objects are consistent with unit covering fraction we omit that possibility here.

\subsection{Galaxy Groups}

\subsubsection{NGC6482}

NGC6482 is the nearest fossil group at a Hubble distance of 55.5 Mpc.  Chandra observations of the group are reported by \citep{Khosroshahi04}.  They find a high central gas density leading to a central cooling time of about $10^8\yr$. The temperature rises inwards from about 0.5 keV at 30 kpc to 0.7 keV near the centre (see also \citep{Kim20}). They find the rise to be consistent with a cooling flow of $2\Msunpyr$ in a steep potential, such as that inferred from the high measured  concentration parameter (\cite{Navarro95}; see also \cite{Buote17}). Their spectral fit is improved if they add photoelectric absorption by a column density of about $2\times 10^{21}\psqcm$. This result is very similar to our RGS solution (Fig. 1), which assumes that the temperature drops at the centre.

\subsubsection{IC4296}

IC4296 is an optically-bright giant elliptical galaxy hosting a powerful twin-jetted radio source PKS 1333-33 (see \citep{Condon21} for spectacular MEERKAT images). It lies at a distance of 47 Mpc and its hot gas has a short central cooling time of about 20 Myr \citep{Lakhchaura2018, Grossova19} have studied the radio source and hot atmosphere of the galaxy including RGS data revealing an unabsorbed cooling flow of $4.5\pm1.0\Msunpyr$, depending on abundance. Our solution, which includes absorption, is similar (Fig. 2). 

We include a power-law spectrum in the spectral fit to account for the nucleus emission reported from Chandra by \citep{Pellegrini03}. It is represented in Fig. 4 by a green line. Interestingly they report an intrinsic absorption column density to the nucleus of $1.1^{+0.8}_{-0.5}\times 10^{22} \pcmsq$, consistent with the intrinsic absorption we measure at $0.8\times 10^{22} \pcmsq$. 

\subsubsection{NGC4325}

NGC4325 is a radio-quiet cool core group/cluster at a distance of 120 Mpc. It has been studied with Chandra and ROSAT by \citep{Russell07}. They found a steep entropy profile in its cool core (see also \cite{Panagoulia2014}) and \cite{McD12} show spectacular H{$\alpha$} filamentation there. It  features in the study of X-ray line broadening by \citep{Pinto2016}. We find an intrinsic column density of $2\times 10^{22} \pcmsq$ and a mass cooling rate of more than $30\Msunpyr$ (Fig 5). 

\subsubsection{NGC3402/3411}

NGC3402 is also known as NGC3411. It has been studied with Chandra and XMM by \citep{Osullivan07}. They find a cool shell of gas at about 20 kpc around a denser core within which the radiative cooling time drops below $10^8\yr$. We find an intrinsic column density of $2.9\times 10^{22} \pcmsq$ and a mass cooling rate of more than $14\Msunpyr$ (Fig 6). 

\subsubsection{NGC4261}
NGC4261 hosts the radio source 3C270, the jets of which are resolved by Chandra (\cite{Worrall10}, \cite{Gliozzi03A&A}. \cite{Zezas05}) have revealed that the nucleus is partially covered by high intrinsic absorption of several times $10^{22}\psqcm$. Line broadening in the RGS spectrum has been studied by \citep{Pinto2016}. We find an intrinsic column density of $2\times 10^{22} \pcmsq$ and a mass cooling rate of  $3.8\Msunpyr$ (Fig 7).

\subsubsection{NGC2300}
Chandra imaging of NGC2300 shows a peaked surface brightness distribution characteristic of a cool core \citep{Grossova2022}. X-ray analysis \citep{Lakhchaura2018} reveals that the innermost gas has a cooling time of less than $2\times 10^8\yr.$ We find an intrinsic column density of $0.15\times 10^{22} \pcmsq$ and a mass cooling rate of more than $0.36\Msunpyr$ (Fig 8).

\subsection{Spiral Galaxies}

\subsubsection{M104}

M104 is also known as NGC4594 or the Sombrero Galaxy. The latter name is due to its appearance: an edge-on disk (SA) galaxy with a thick dust lane. Chandra and XMM observations of it have been reported by \cite{Li11}, \cite{Li07}.  Many compact sources, presumably low X-ray binary systems (LMXB etc., see discussion in \cite{Ivey24}) as well as a point-like nucleus are seen, together with some diffuse emission. 

We are only concerned with the diffuse component and model the point sources as collectively producing a simple hard bremsstrahlung spectrum shown in green. This includes the central power-law emission reported from Chandra by \citep{Pellegrini03a}.  The strong Fe XVII  and OVIII emission lines are presumed to originate in the diffuse component. This is modelled as a Hidden Cooling Flow, for which we determine a rate of $\sim0.3\Msunpyr$ with an intrinsic column density of $0.5\times 10^{22} \pcmsq$ (Fig 9).

\subsubsection{M51}

The nearby galaxy M51 at a distance of 8,5 Mpc is also known as the Whirlpool Galaxy or NGC5194, It is interacting with a smaller companion galaxy NGC5295. Its X-ray emission was studied using XMM Epic by \citep{Owen09}. They report many point sources and diffuse emission.  

\citet{Zhang22} used the XMM RGS data to determine the nature of its soft X-ray spectrum and manage to roughly map its origin. They also find evidence for strong intercombination and forbidden OVII emission at about 22\AA\,  which is a marker for Charge Exchange emission due to interaction between the hot and cold gas components. This OVII component is clearly not fitted in our RGS spectrum and is not part of our modeled  cooling flow, which is  at a level of about $1.1 \Msunpyr$ with an intrinsic column density of $>0.8\times 10^{22} \pcmsq$ (Fig 10). 

\subsubsection{NGC253}

NGC253 is a nearby (3.2 Mpc) starburst galaxy, also known as the Sculptor galaxy.  It is oriented almost edge-on at 78 deg. It has been studied with XMM by \citep{Bauer07} and deeply with Chandra, most recently by \cite{Lopez23}.  Outflows, short cooling times, and bulk radiative  cooling are observed or inferred. The X-ray brightest region is less than 1 arcmin across and centred on the nucleus.  A bremsstrahlung continuum is included in the spectral fit to account for LMXB.
we determine a cooling rate of $\sim0.4\Msunpyr$ with an intrinsic column density of $2.5\times 10^{22} \pcmsq$ (Fig 11).

\begin{figure}
    \centering    
\includegraphics[width=0.48\textwidth]{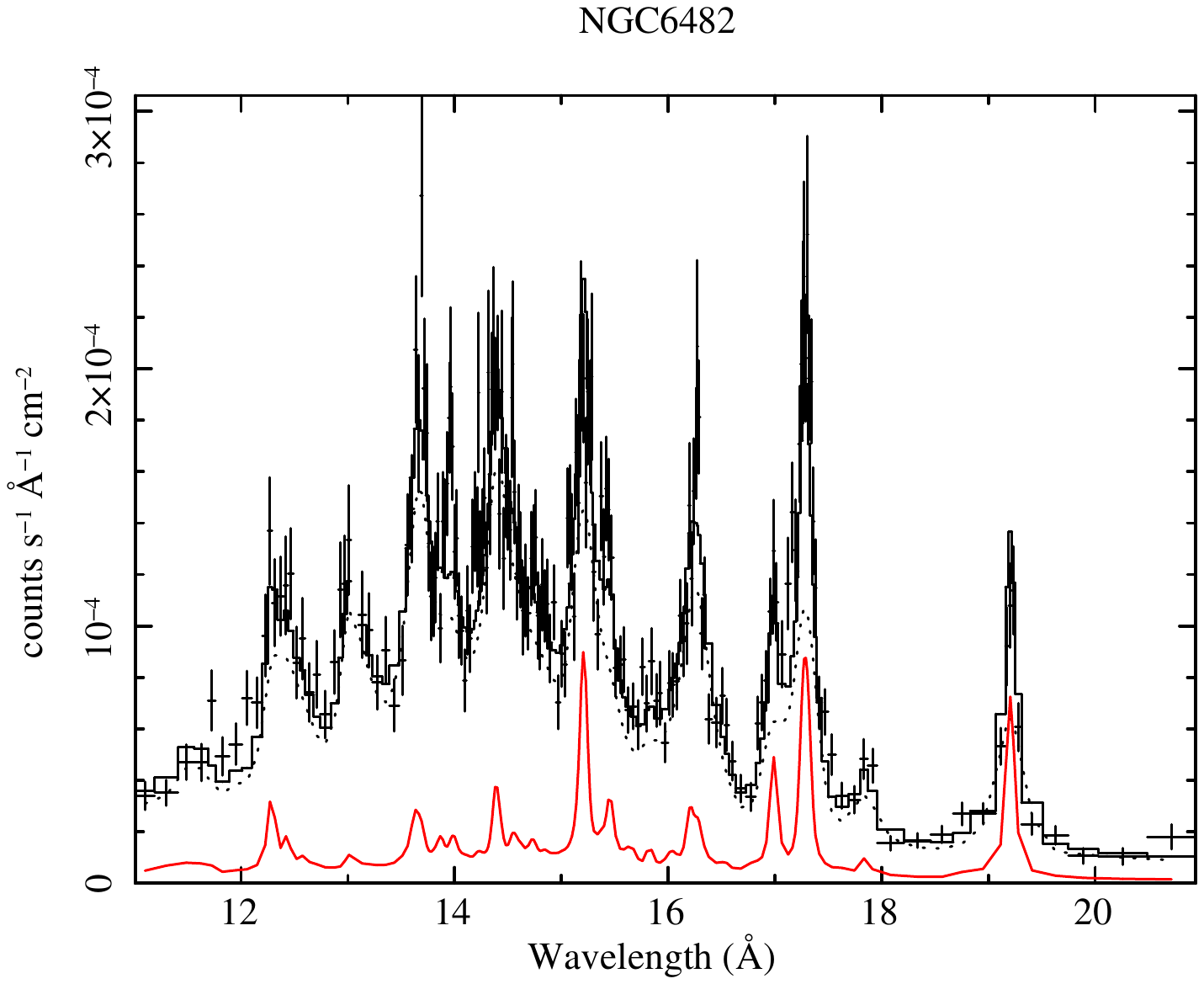} 
\includegraphics[width=0.48\textwidth]{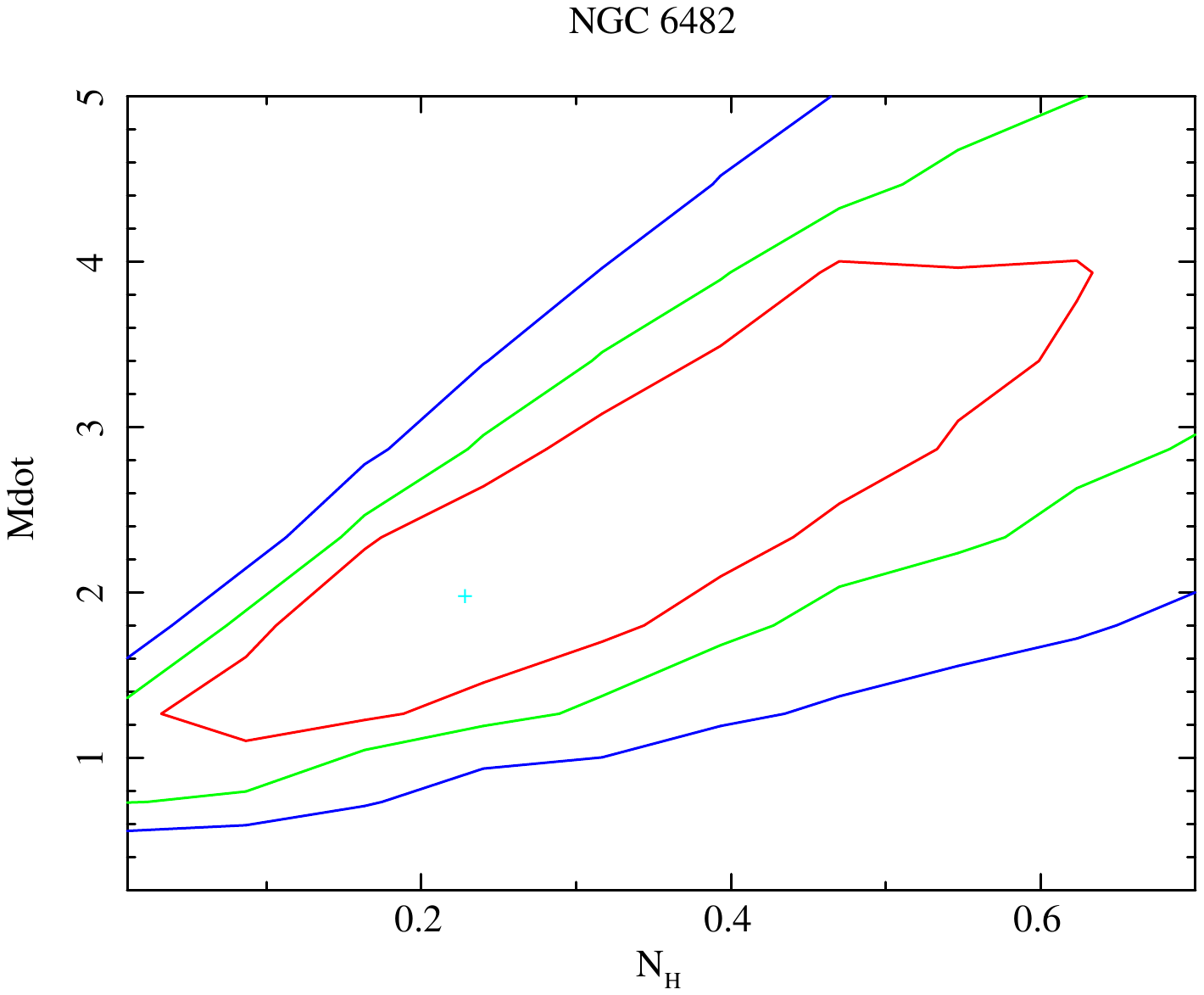}
    \caption{ a) RGS spectrum and b) hidden mass cooling rate versus total interleaved column density. } 
\end{figure}

\begin{table*}
	\centering
	\caption{Spectral Fitting Results. The units of column density $N_{\rm H}$ are $10^{22}\pcmsq$,  $N_{\rm H}$ is the Galactic column density and $N_{\rm H}'$ is  the intrinsic value. The temperature $kT$ of the surrounding gas modelled by \textsc{apec} and maximum of \textsc{mkcflow} component $kT$ (which are the same) is in $\keV$, $Z$ is  abundance  relative to Solar and $\dot M$ is in $\Msunpyr$.  (f) means that a parameter is fixed. The uncertainty contours for $\dot M$ and $N_{\rm{H}}$ are often unclosed at higher values and we therefore give a  90\% confidence upper limit for the mass flow rate.}
	\begin{tabular}{lcccccccccr} 
		\hline
		Cluster & $N_{\rm H}$ &  $kT$ & $Z$ &$z$ & $Norm$ &   ${N_{\rm H}}^{'}$ &  $\dot M$ & Limit &$\chi^2/{\rm dof}$ &\\
		\hline
		 & $10^{22}\cmsq$ & $\keV$ & $Z_{\odot}$ & & &  $10^{22}\cmsq$ & $\Msunpyr$ & & $\Msunpyr$ \\
		 \hline
   NGC6482 & 8e-2& 0.79 & 0.35 &1.26e-2 &1.1e-3  & $0.22$& $1.84$ & $>0.97$ & 545/446  \\
   IC4296& 3.3e-2 & 0.88& 0.2 & 1.2e-2 & 4.8e-4 & 0.82 &$3.8$& $>1.3$ &144/129  \\
   NGC4325& 2.7e-2 & 0.96& 0.22 &2.6e-2 &3.5e-3 &  2 & 63 &$>30$ &158/146 \\
   NGC3402 & 4.7e-2 & 0.95 & 0.36 &1.5e-2&3.2e-3 &  2.9& 14 &>3 &1017/997  \\
   NGC4261 &1.6e-2& 0.81 & 0.13 &6.9e-3 & 8.1e-4 &2.32 & 3.8 &>0.7  & 320/298 \\
   NGC2300 &5.8-2&0.79&0.18 & 7e-3&7.6e-4&0.15& 0.36& >0.15& 100/100 \\
   M104 &3.7e-2 & 0.71 & 0.2f & 4e-3 & 2.6e-6& 0.5 &0.3 &--& 387/364  \\
   M51  & 6.8e-2 & 0.307 &0.2f & 2.1e-3 & 2.5e-4 & >0.8 & 1.1 & -- & 1346/1008 \\
   NGC253 &0.24  & 0.64 & 0.2f & 1.3e-7 & 2.23e-2 &  2.5 &  0.4 & -- &
   1056/855   \\
		\hline
	\end{tabular}
\end{table*}

\section{The Efficiency of Feedback}

The 6 new groups examined here show Hidden Cooling Flows of a few $\Msunpyr$, similar to the 8 examined earlier. NGC4325 is the exception with at least $30\Msunpyr$ being required. It is however the most distant group we have studied and has  a luminosity comparable to a poor cluster. Our sample of groups is by no means complete but indicates that HCF are common in groups which have a central elliptical galaxy. 

Indeed, if we include our observations of isolated and cluster elipticals, it seems that all massive elliptical galaxies may host an HCF. A minimum mass inflow rate may be that due to stellar mass-loss within the galaxy. Radio mode feedback involving jets does not appear to prevent  the accumulation of  cooling gas in their centres, with the jetted galaxies NGC3516 (Fornax A), IC4296 (PKS1333), NGC4261 (3C270), M84 (3c272.1) let alone Cygnus A shown in HCFIV  hosting clear examples of HCF. Of course, the heating that jets do provide is concentrated at their ends  not near the centre where most of the cooling gas resides. 

As we showed in HCFIV \cite{Fabian24b}, radio mode feedback is rarely 100 per cent efficient\footnote{It is labelled "Radiative Efficiency" in Fig.9 of that paper, but should really be "Feedback Efficiency"}. Defining the efficiency as 
$(1-{\rm HCR}/{\rm IMR})$, where HCR is the inner Hidden rate from the RGS and IMR is the Chandra X-ray imaging rate dominated by the outer regions and reported by \cite{McDonald2018}, we found values of 40 to 95 per cent for luminous clusters where IMR exceeds $10\Msunpyr$. However where it is less. i.e. the groups and ellipticals, then a general result is that HCR approximately equals IMR. The objects here are the groups and elliptical galaxies where any feedback is not suppressing central cooling.

\begin{figure}
\centering  
\includegraphics[width=0.48\textwidth]{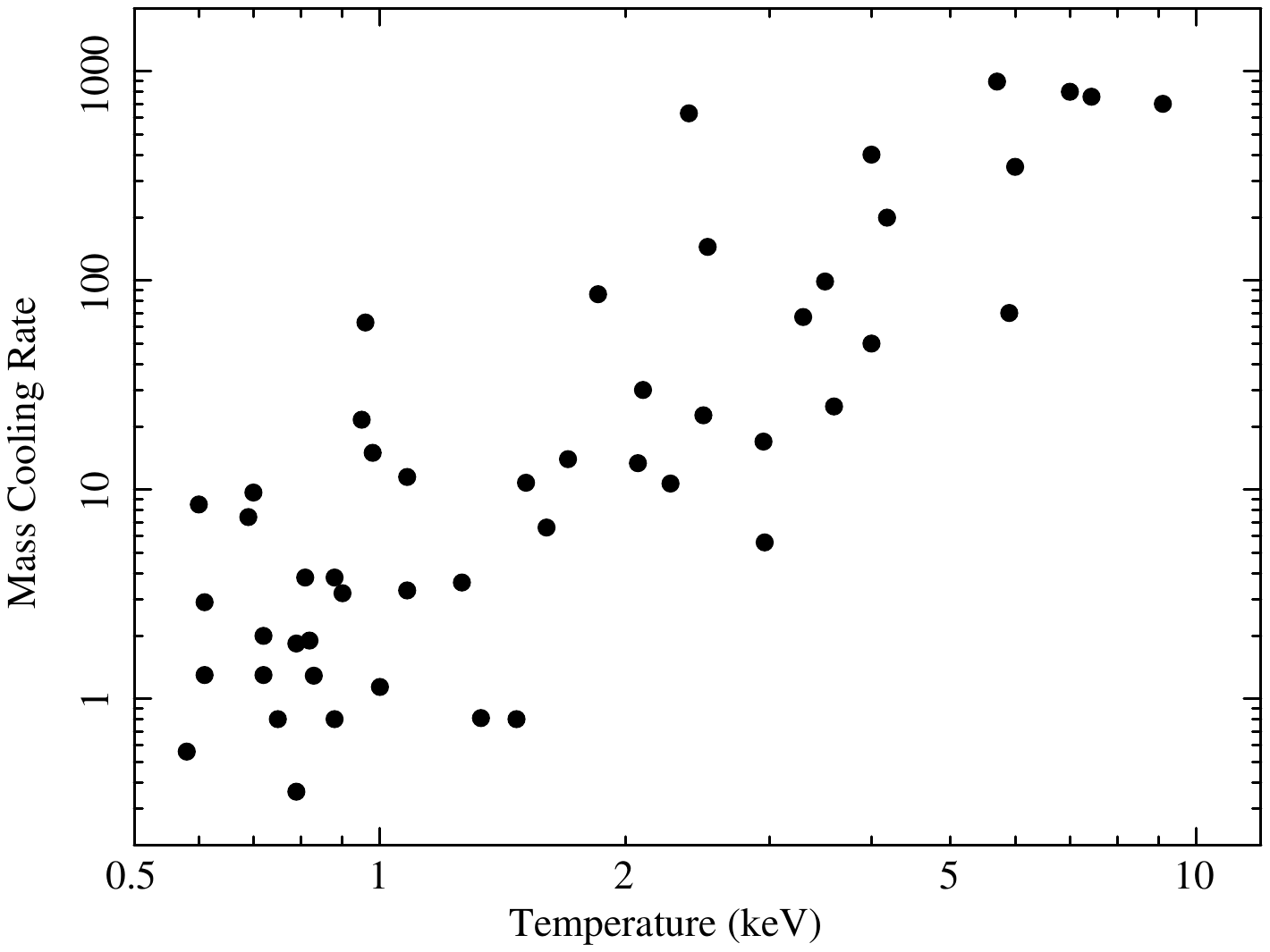}   
    \caption{ Hidden/absorbed mass cooling rate (from central gas) plotted against surrounding (outer gas) temperature of our whole sample (Table 2) of 50 ellipticals, groups and  clusters. }
\end{figure}

\begin{figure}
\centering  
\includegraphics[width=0.48\textwidth]{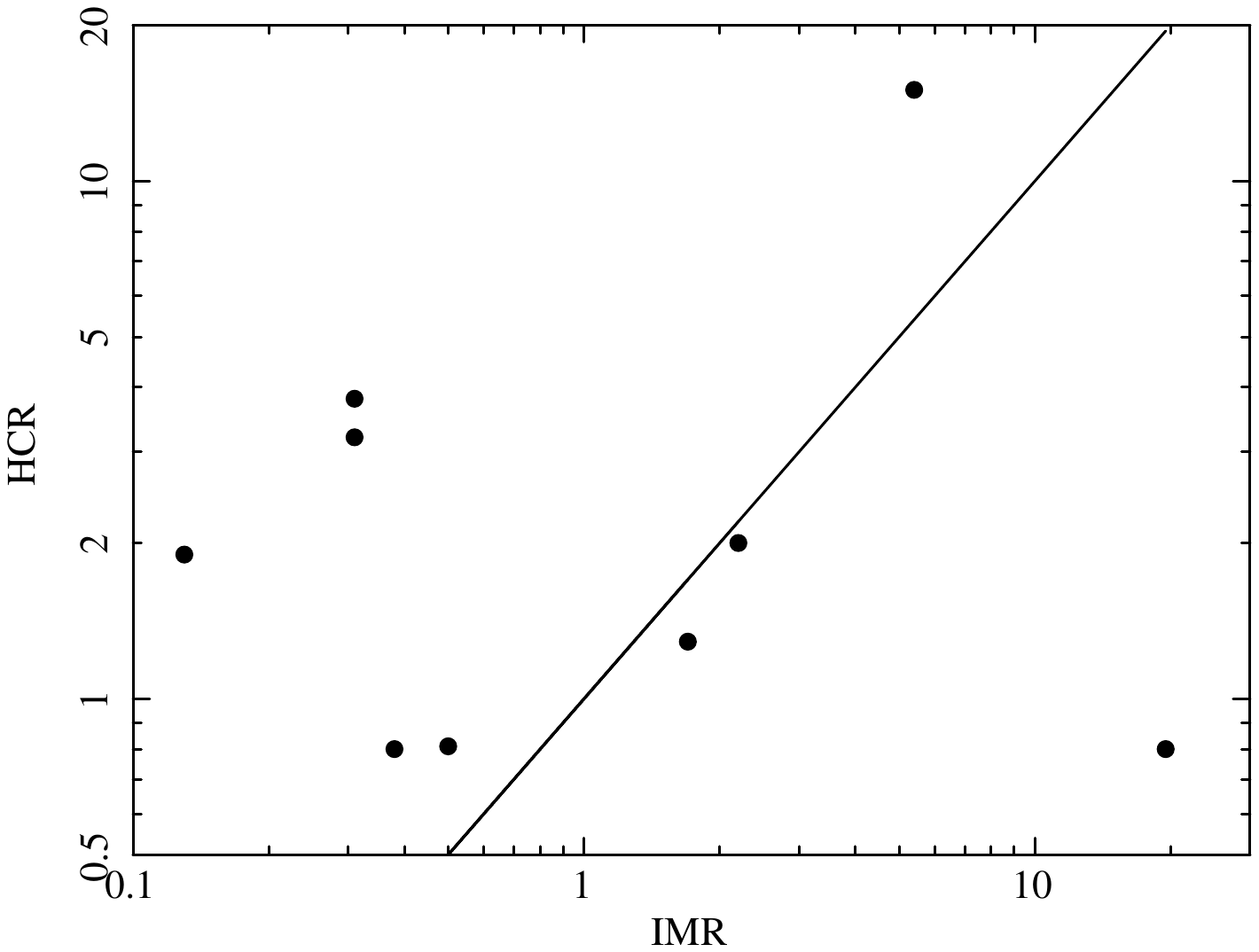}   
    \caption{ Hidden mass cooling rate (HCR) plotted against imaging mass cooling rate (IMR) from \citep{McDonald2018} of ellipticals, groups and clusters studied in HCFI to HCFV (Table 2), where the outer temperature $kT< 1\keV$. Objects close to and above the line have   HCR $\gtrapprox$
    IMR, so feedback has not greatly suppressed cooling. The object on the lower right well below the line is M87 where the hidden cooling rate (small scale) is $0.8\Msunpyr$ and the large scale imaging rate is $19.5\Msunpyr$. Jetted AGN feedback has diminished  the inner cooling flow but not eliminated it. Points from left to right are M84, (up) NGC4261, NGC1316, NGC4636, NGC1600, NGC5846, NGC5813, HCG62 and M87.} 
    
\end{figure}
\section{Low-mass star formation}

A major problem for Hidden Cooling Flows is what happens to the cooled material. As discussed in the previous HCF papers we have identified low mass stars, principally brown dwarfs, as the fate of much of the cooled material. Star formation in the deepest extended gravitational potentials in the Universe, the centres of massive elliptical galaxies is special. Generally there is no rotational support for cooled gas, the velocities of stars or other collapsed objects are several $100\kmps$ and the stellar density very high. If brown dwarfs of typical mass $0.03\Msun$ are accumulating there then the number density must exceed $10^3\ppccu$. 

Whether such brown dwarfs can grow more massive depends on how well they can accrete further from their surrounding natal gas \citep{Krumholz16}. The accretion radius for a brown dwarf  travelling at $100\kmps$ is less than $10^{11}\cm$. So it will not accrete from general gas in the galaxy core but it might be able to accrete from its birth cloud if it is comoving. The result  depends on the properties, including geometry, of that cloud material and how that is being pushed and pulled around in the deep central potential well. If in threads of thin magnetised filaments as seen further out in cool core clusters e.g. from their H$\alpha$ emission, then there may be little comoving gas for new brown dwarfs to grow larger. 

The existence of a bottom-heavy initial-mass function in the cores of massive elliptical galaxies has been established by optical observations of such regions by e.g.  \cite{vDC2010, VanDokkum24, Gu22, Oldham2018}. Hidden cooling flows can provide a means to grow such a region, but the detailed pathway remains unclear.

\section{Spiral Galaxies}

We have noticed that the RGS spectra of many bright nearby spiral galaxies resemble those of hidden cooling flows due to the presence of sub-keV X-ray emitting gas. Such gas probably  constitutes the CircumGalactic Medium. In the words of \cite{Tumlinson17ARA&A..55..389T}, the CGM
is a source for a galaxy’s star-forming fuel, the venue for galactic feedback and recycling, and perhaps the key regulator of the galactic gas supply. As mentioned in the Introduction (see e.g. \cite{Sultan25}), several recent studies propose that cooling flows may be operating in the CGM. 

Here we take a simplistic approach of applying our basic HCF model to the RGS spectra of 3 nearby spiral galaxies, the Sombrero Galaxy (M104), the Whirlpool Galaxy (M51) and the Sculptor Galaxy (NGC253). We include a bremsstrahlung continuum to mimic the spectra of point sources (LMXB, Low Msss X-ray Binaries etc) and nucleus and adopt a metallicity fixed  at 0.2 Solar. The model thus consists of a uniform hot medium  plus a hidden (absorbed) flow cooling from that upper temperature and a bremsstrahlung component. The quality of the fits in terms of $\chi^2/{\rm dof}$ is poor (1.06, 1.33 and 1.24 respectively) compared with the groups studied here where the $\chi^2/{\rm dof}$ ranges from $1-1.22$ with a mean of  1.17. Note that a significant deviation in the M51 spectrum making it the worst fit, occurs in the OVII emission at 22\AA\ due to the unmodelled charge exchange component (see \cite{Zhang22}. The mass cooling rates range from $0.3-1.1\Msunpyr$ and the intrinsic absorbing column densities range from $0.5-2.5 \times 10^{22} \pcmsq$. Most of the line emission is from the cooling flow component. The mass cooling rate is then inversely proportional to the assumed abundance (fixed at 0.2 here).  

We conclude here that absorbed cooling flow models may indeed be appropriate for star-forming  spiral galaxies. 
The geometry of the flow is likely to be different (at least 2 dimensional) compared with that in elliptical galaxies and groups where it could be more one-dimensional. There is also likely to be centrifugal support for gas in the spirals. 

\section{Discussion}

We have found Hidden Cooling Flows in 6 more galaxy groups. The mass cooling rate is typically a few $\Msunpyr$ but can rise tenfold where there is a higher outer reservoir temperature. The intrinsic column densities range from $0.15-2.9\times 10^{22}\pcmsq$. We also find evidence that modest absorbed cooling flows exist in Spiral Galaxies. 

AGN Feedback appears to operate well on large kpc scales in many clusters but is unable to suppress radiative cooling on smaller scales in their centres as well as in most groups and elliptical galaxies.  This is an obvious  consequence of jets being highly directional and able to push through the central regions before releasing most of their power outside the form of in bubbles and sound waves etc.

The outstanding question of what happens to the cooled gas remains open. If it cools all the way down close to zero then low-mass stars is an option (see section above and \cite{Fabian24}).  As discussed in HCFIV \citep{Fabian24b},
gas can be detected at temperatures around $3\times 10^5\K$ through the MIR 7.65${\mu}$m emission line of NeVI. This is in principle detectable in bright objects with JWST. 

If no such NeVI emission is seen then it  is possible that the gas cooling below $10^6\K$ (the lower X-ray spectral limit in our models) has mixed into the surrounding cooler gas which then cool together  Either that or heating has somehow implausibly sought out the coolest part of the flow where  the radiative cooling rate is fastest.

High resolution X-ray spectroscopy of the cooling gas in galaxies, groups  and clusters can continue to be obtained with the XMM RGS and from XRISM \citep{XRISM2018}, if the Gate Valve is opened.  Future spectroscopy at yet higher resolution both spectrally and spatially relies on NewAthena.

\begin{table*}
	\centering
	\caption{The total sample of 50 objects studied so far. The (HCF) mass cooling rates $\dot M$ for Perseus, A1795, and Phoenix should be treated as lower limits. }
	\begin{tabular}{lcclcc} 
		\hline
        &kT<1.5 keV & & &kT>1.5 keV &\\
        \hline
        Name & $\dot M$ &  $kT$ & Name & $\dot M$ & $ kT$ \\
         & $\Msunpyr$ & keV & & $\Msunpyr$ & keV \\
         \hline
         NGC4552 &0.56 &0.58 & A3581 & 10.8 & 1.51 \\
         NGC 720 & 1.3 & 0.61 & A262 & 6.6 & 1.6\\
         NGC1332 & 2.9 & 0.61 & Cen & 15 & 1.7 \\
         NGC1404 & 7.4 & 0.69 & 2A0335 & 86 & 1.85 \\
         MRK1216 & 9.7 & 0.7 & A2052 & 13.4 & 2.07 \\ 
         IC1459 & 1.3 & 9,72 & RXJ0821 & 20 & 2.1  \\ 
         NGC5831 & 2 & 0.72 & S159 & 10.7 & 2.27 \\ 
         NGC4636 & 0.8 & 0.75 & A1068 & 630 & 2.39 \\
         NGC2300 & 0.36 & 0.79 & A496 & 22.7 & 2.49 \\
         NGC5846 &1.29 & 0.83 & A1664 & 145 & 2.52 \\
         NGC6482 &  1.47 & 0.79 & Hydra A & 17 & 2.95 \\
         NGC4261 & 3.8 &  0.81 &  A2199 5.6 & 2.96 \\
         M84 & 1.9 & 0.82 &  A2597 & 5.6 & 2.96 \\
         IC4296 & 3.8 & 0.88  & A3112 & 99 & 3.51\\
         NGC4649 & 0.8 & 0.88 & A85 & 25 & 3.6 \\
        NGC1316 & 3.2 & 0.9 & Per & 50 & 4\\
        NGC3091 & 8.5 & 0.91 & RXJ1532 & 400 & 4\\
        NGC5044 & 21.6 & 0.95 & A1795 & 21.6 & 0.95 \\
        NGC4325 & 63 & 0.96 & Zw3146 & 895 & 5.7\\
        HCG62 & 15 & 0.98 & A1835 & 70 & 5.9 \\
        M49 & 1.14 & 1.0 & Cyg A & 350 & 6 \\
        NGC1399 & 3.3 & 7 1.08 & Phoenix & 800 & 7\\
        NGC533  & 11.5 & 1.08 & RXJ1504 & 757 & 7.44\\
        NGC1550 & 3.6 & 1.26 & MACS 1931 & 700 & 9.1\\
        NGC1600 & 0.81 & 1.33 \\
         M87 & 0.8 & 1.47 \\
		\hline
	\end{tabular}
\end{table*}

\begin{table*}
\centering
\caption{Details of the Objects and their RGS exposures}
\begin{tabular}{lcccp{5cm}}
\hline
Object  & RA (de)    & Dec (deg)  & Exposure (ks) & OBSIDs \\ \hline
NGC6482 & $267.9532$ &  $23.0719$ &  $365.1$ & 0304160401  0304160501 0304160601  0304160801  0822340101  0822340201  0822340301 \\
IC4296  & $204.1626$ & $-33.9659$ &  $128.2$ & 0672870101 \\
NGC4325 & $185.7778$ &  $10.6212$ &   $42.6$ & 0108860101 \\
NGC3402 & $162.6102$ & $-12.8446$ &   $51.8$ & 0146510301 \\
NGC4261 & $184.8467$ &   $5.8249$ &  $264.7$ & 0056340101 0502120101 \\
NGC2300 & $113.0834$ &  $85.7090$ &  $104.9$ & 0022340201 \\
M104    & $189.9976$ & $-11.6231$ &  $302.0$ & 0084030101  0900170101 \\
M51     & $202.4842$ &  $47.2306$ & $1701.5$ & 0112840201  0303420101 0677980701  0824450901  0830191501  0852030101  0883550201  0212480801 0303420201  0677980801  0830191401  0830191601  0883550101  0883550301 \\
NGC253  &  $11.8880$ & $-25.2881$ &  $505.7$ & 0125960101  0125960201 0152020101  0304850901  0304851001  0304851101  0304851201  0304851301\\
\hline
\end{tabular} 
\end{table*} 

\begin{figure}
    \centering    
\includegraphics[width=0.48\textwidth]{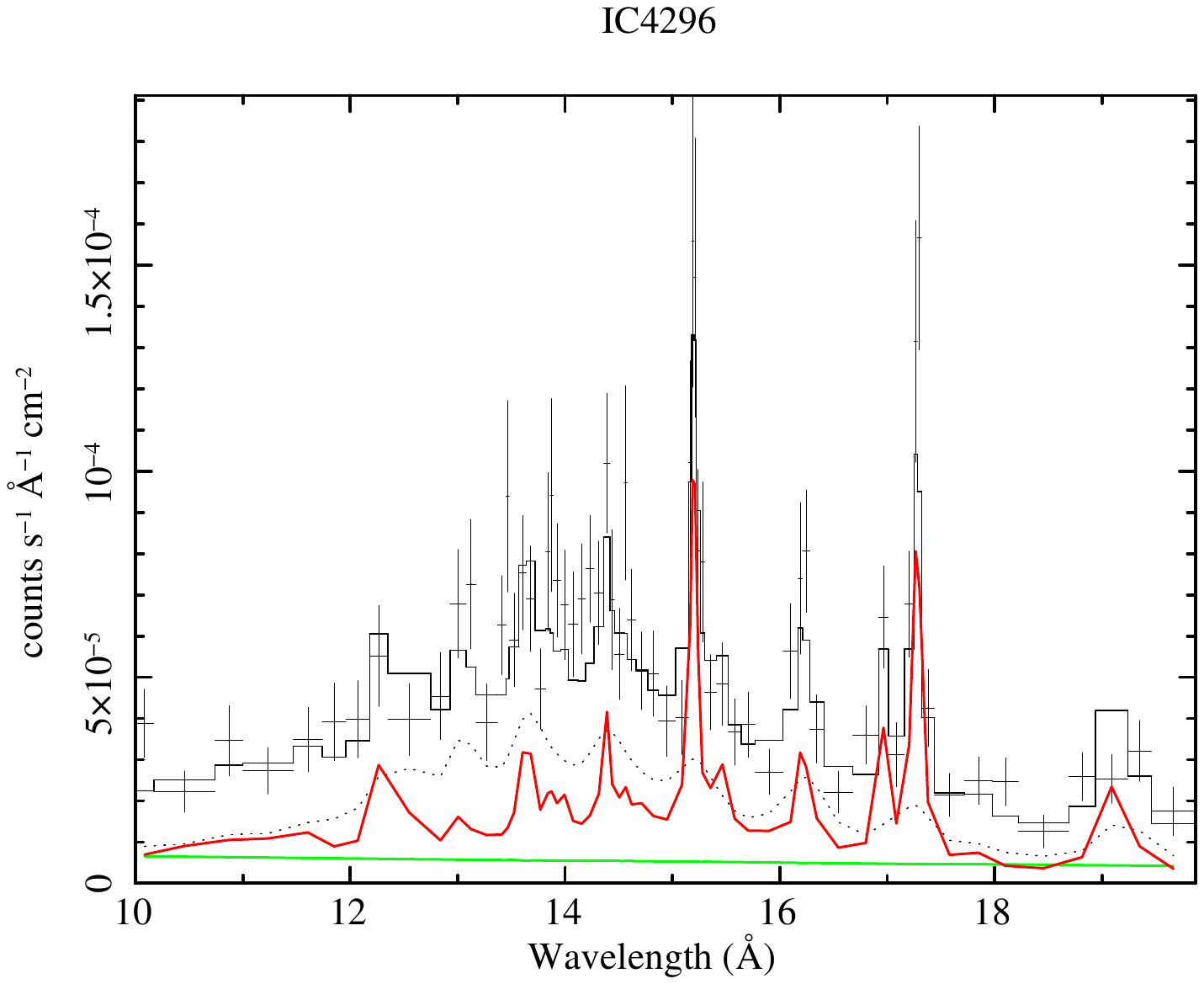} 
\includegraphics[width=0.48\textwidth]{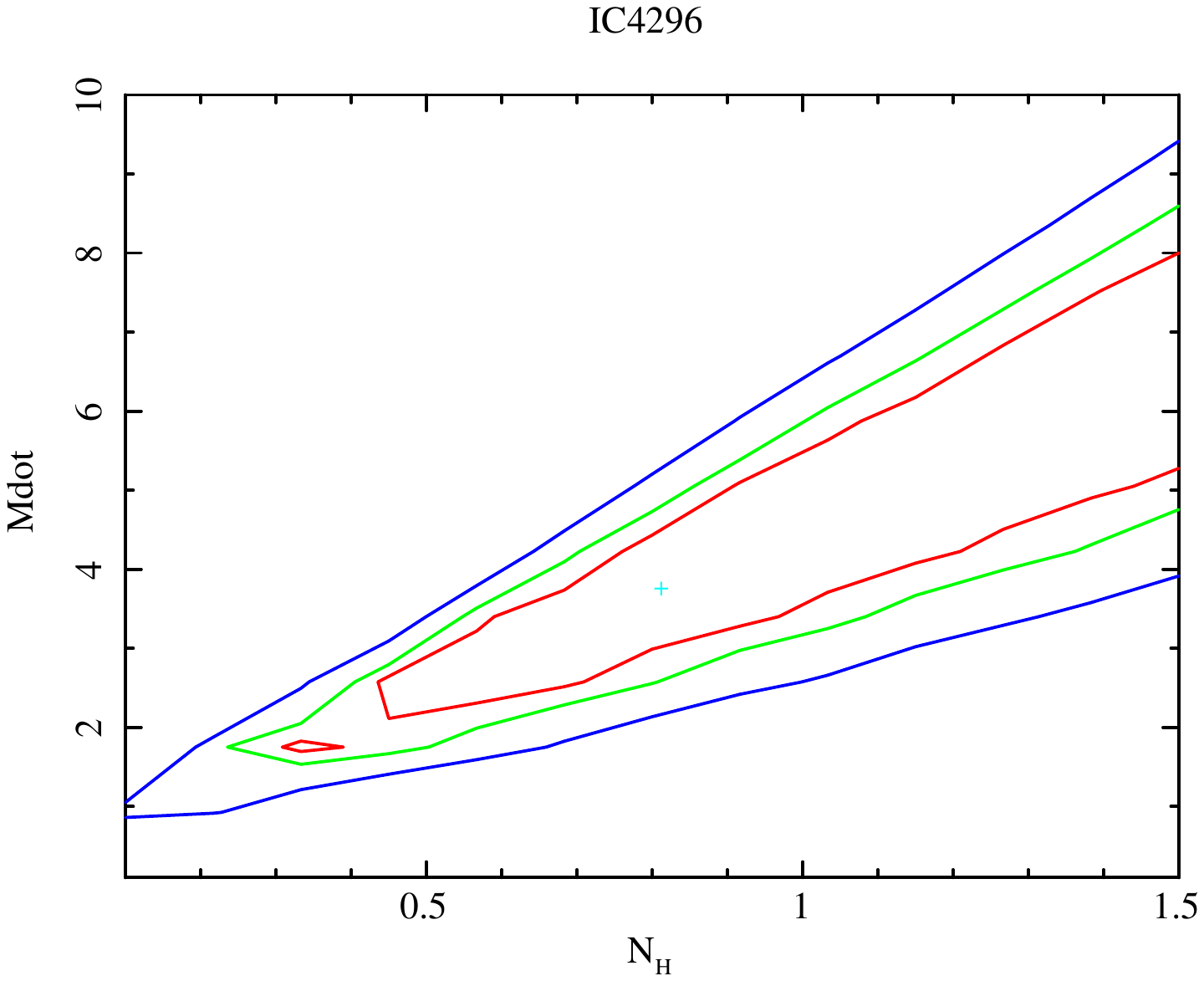}
    \caption{ a) RGS spectrum and b) hidden mass cooling rate versus total interleaved column density. }
\end{figure}

\begin{figure}
    \centering    
\includegraphics[width=0.48\textwidth]{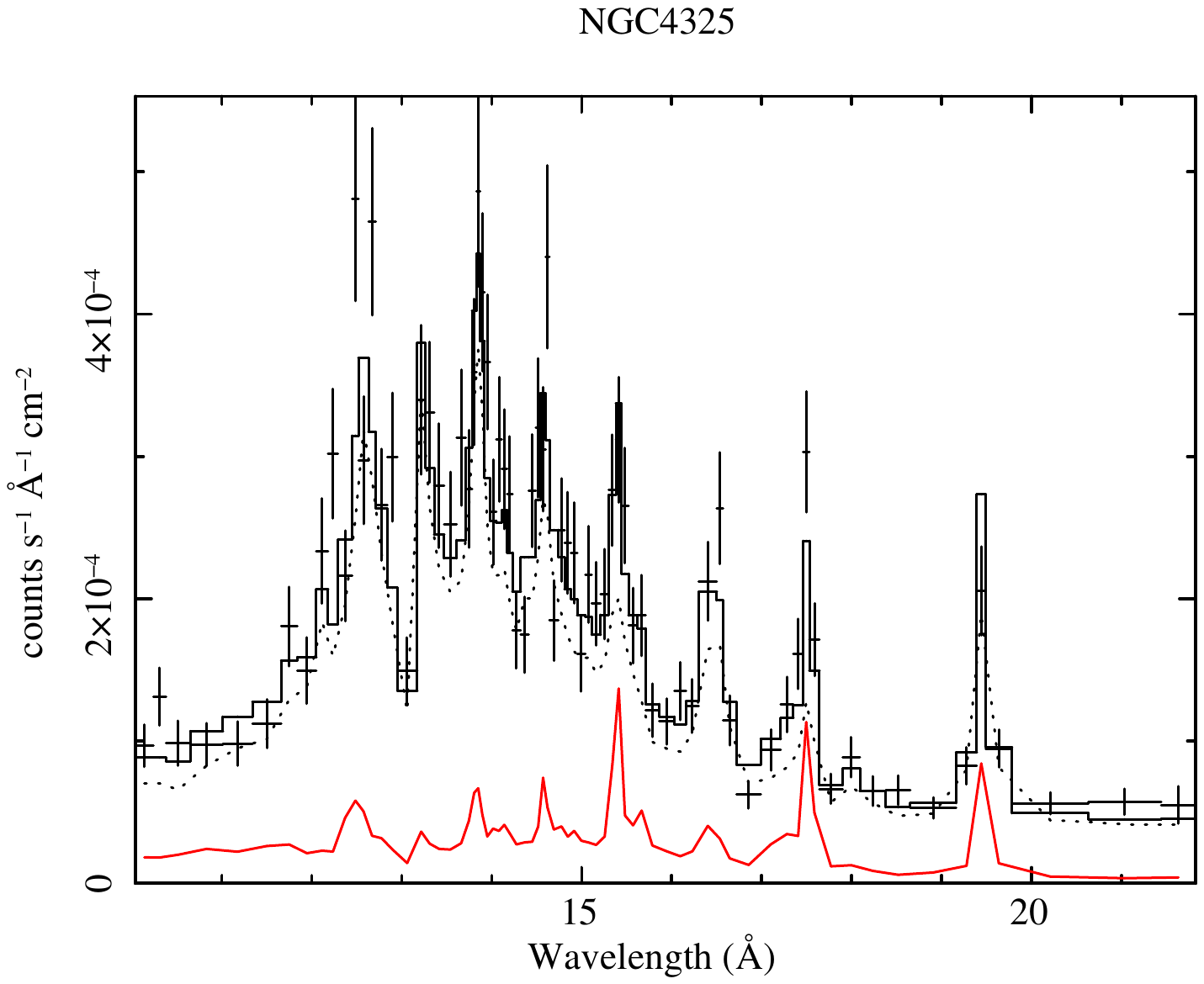} 
\includegraphics[width=0.48\textwidth]{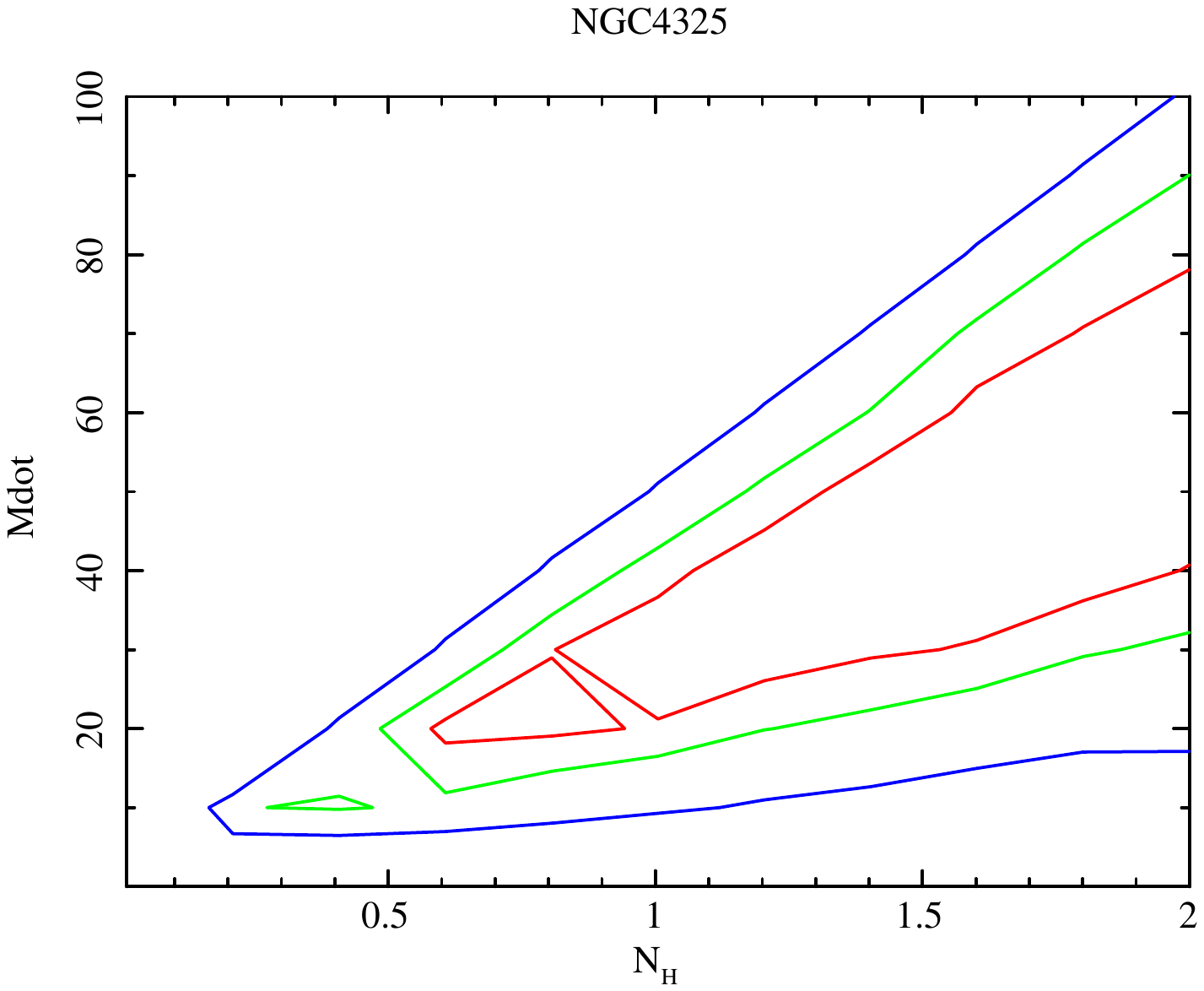}
    \caption{ a) RGS spectrum and b) hidden mass cooling rate versus total interleaved column density. }
\end{figure}

\begin{figure}
    \centering    
\includegraphics[width=0.48\textwidth]{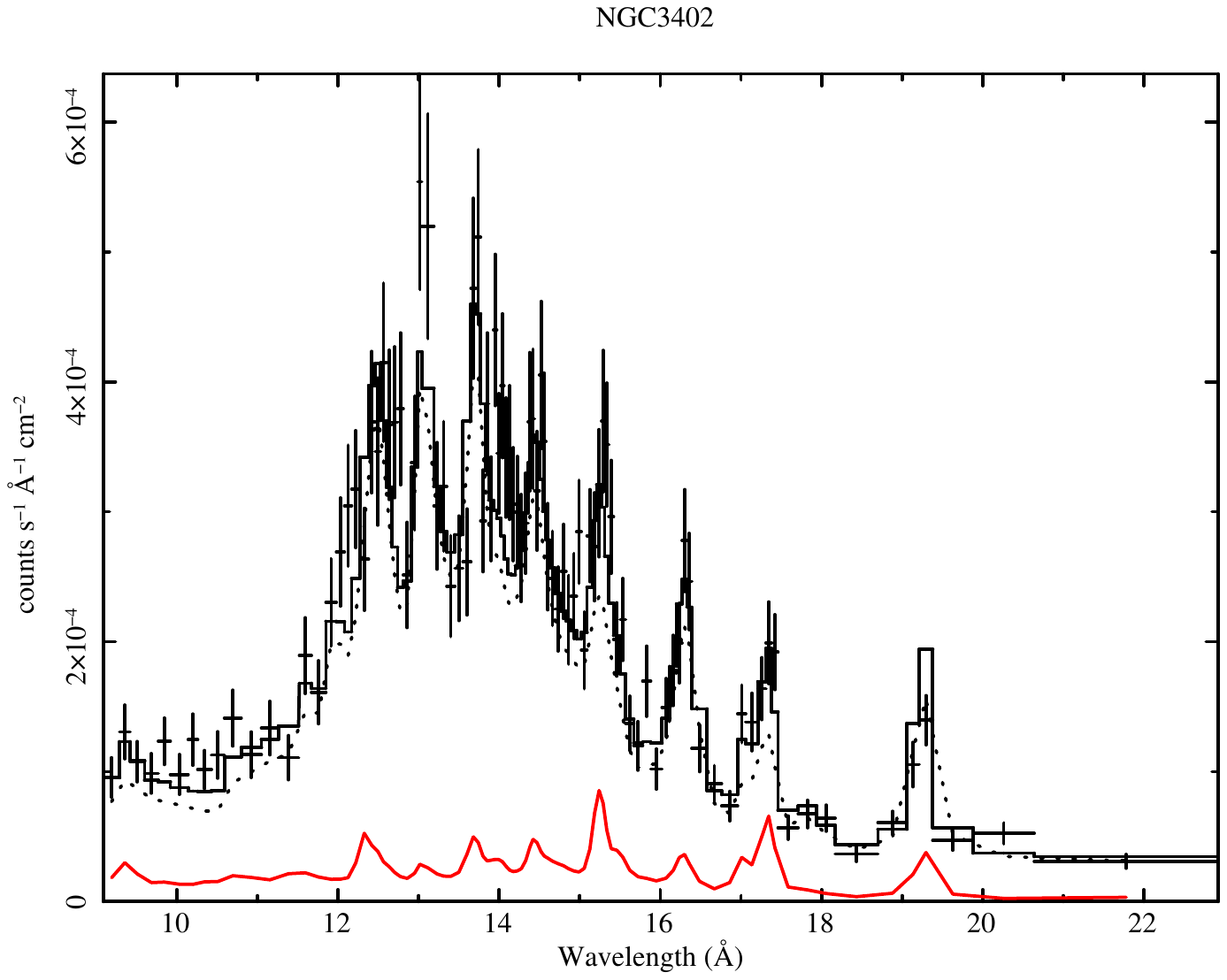} 
\includegraphics[width=0.48\textwidth]{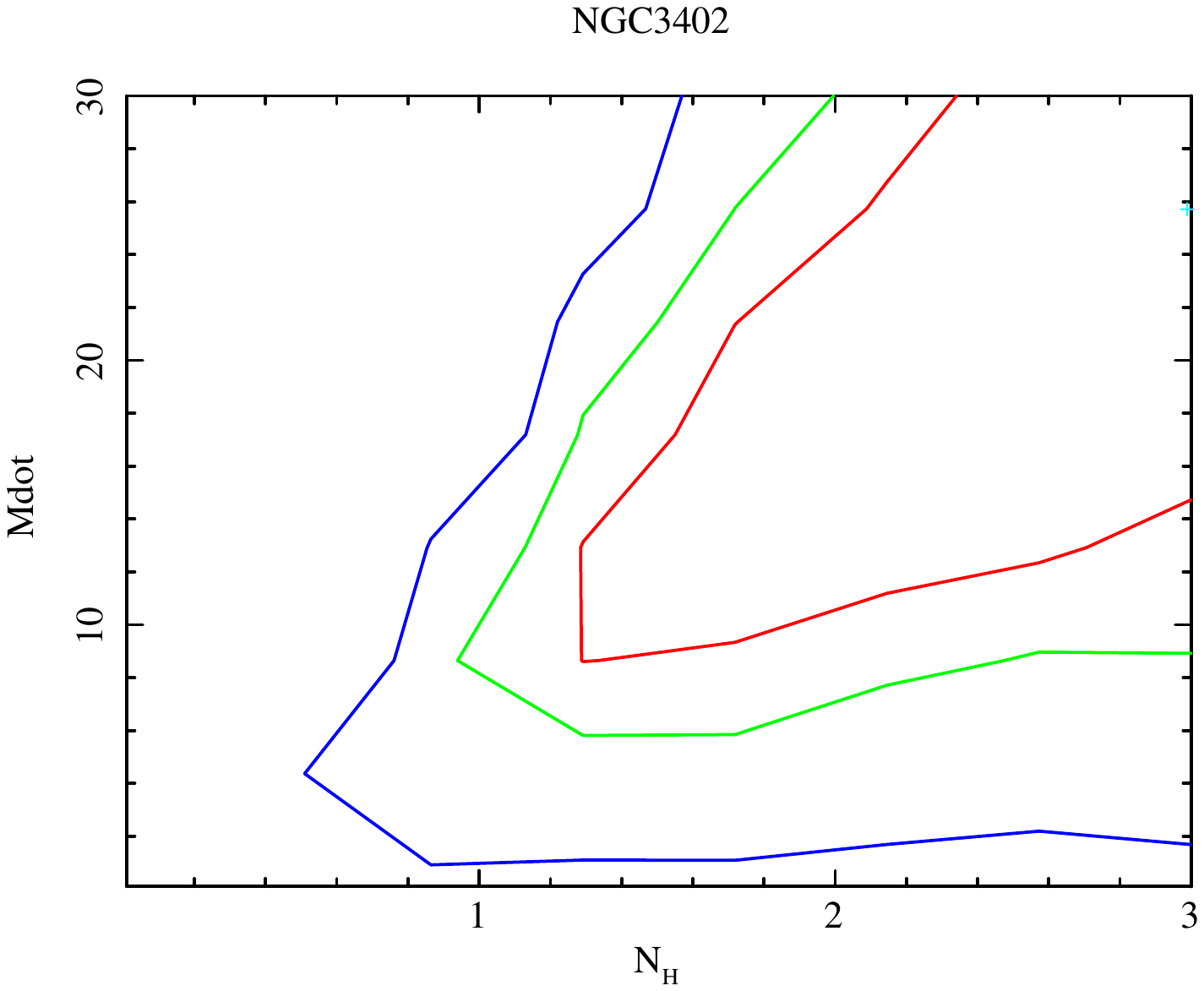}
    \caption{ a) RGS spectrum and b) hidden mass cooling rate versus total interleaved column density. } 
\end{figure}

\begin{figure}
    \centering    
\includegraphics[width=0.48\textwidth]{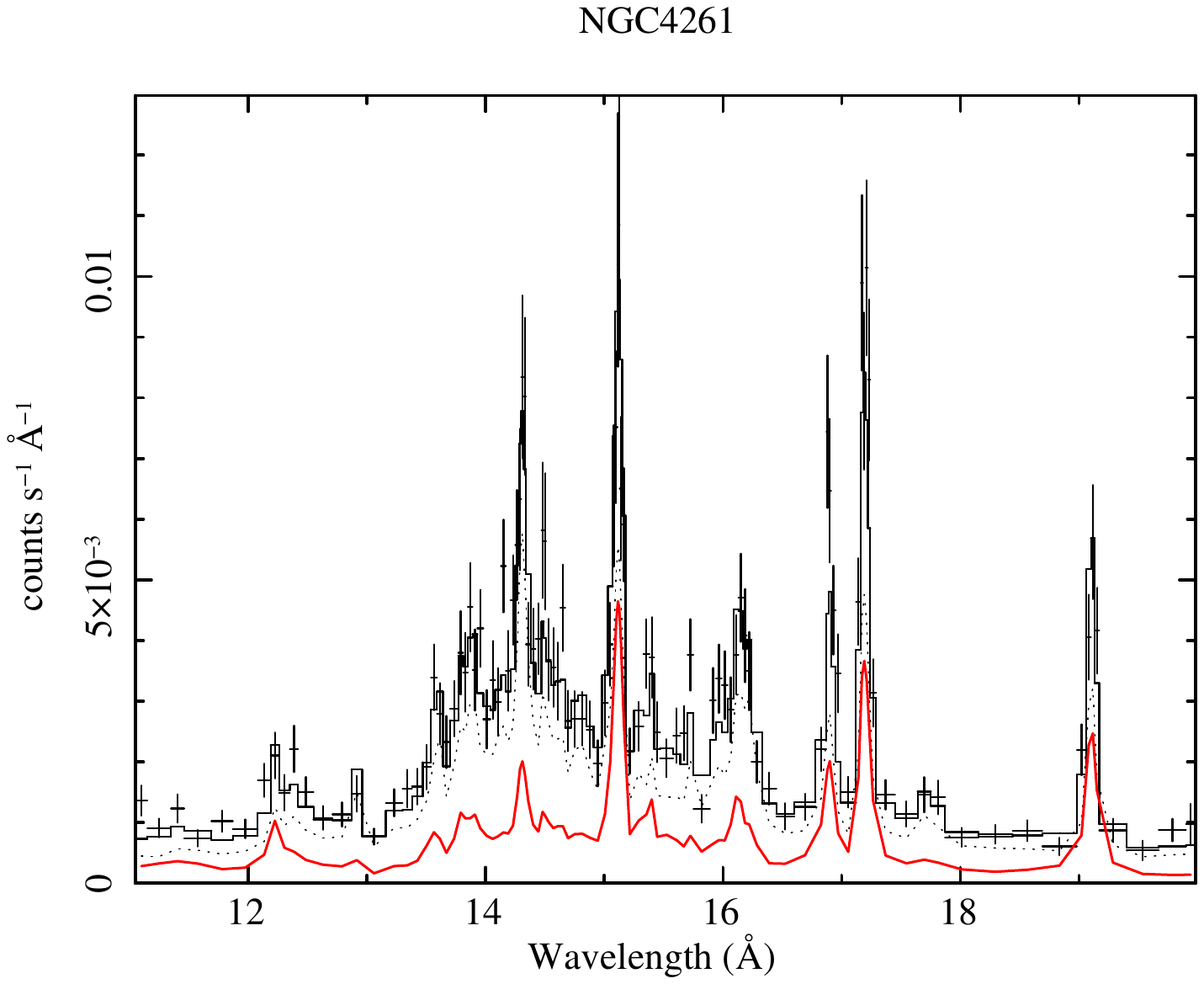} 
\includegraphics[width=0.48\textwidth]{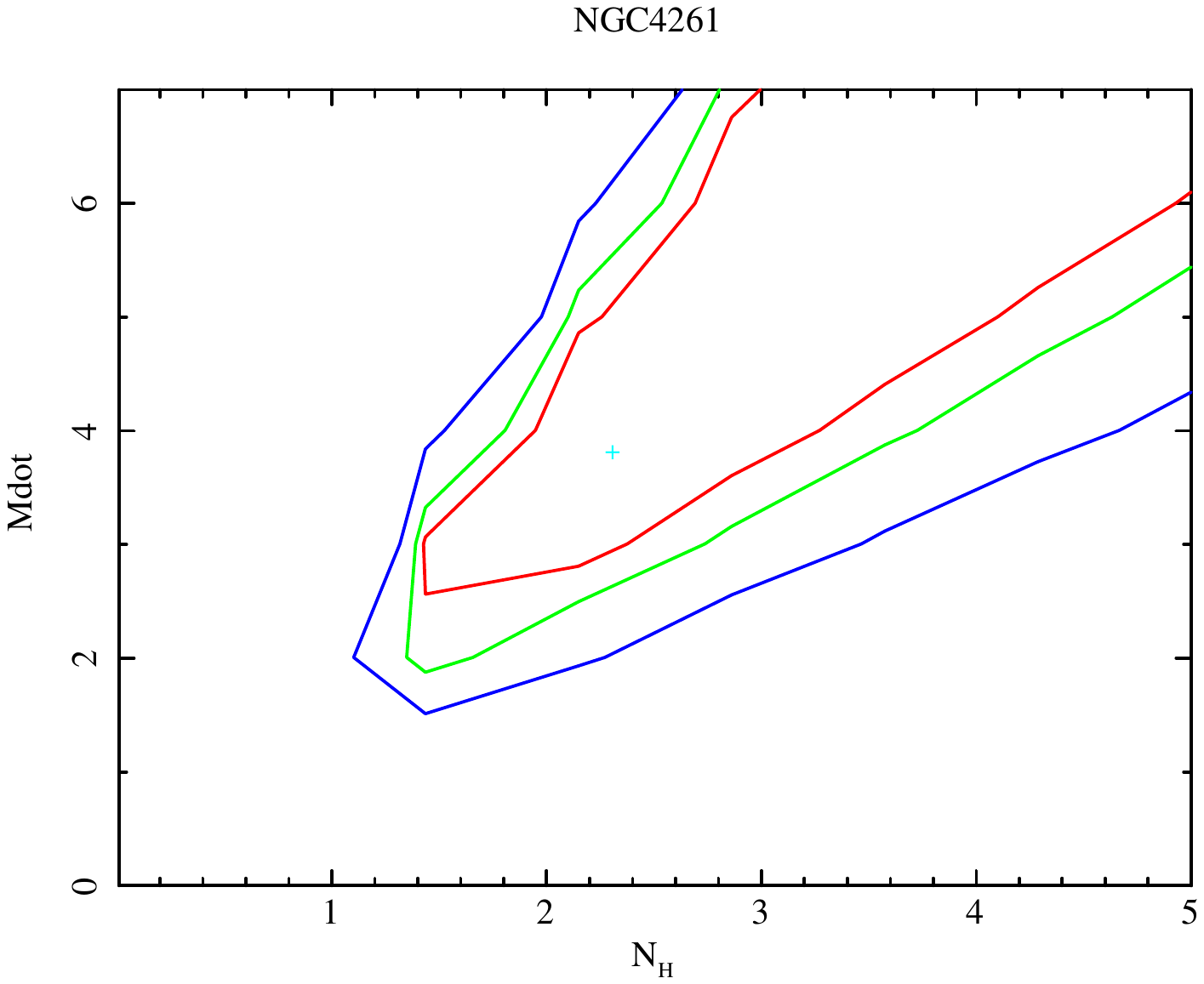}
    \caption{ a) RGS spectrum and b) hidden mass cooling rate versus total interleaved column density. }
\end{figure}

\begin{figure}
    \centering    
\includegraphics[width=0.48\textwidth]{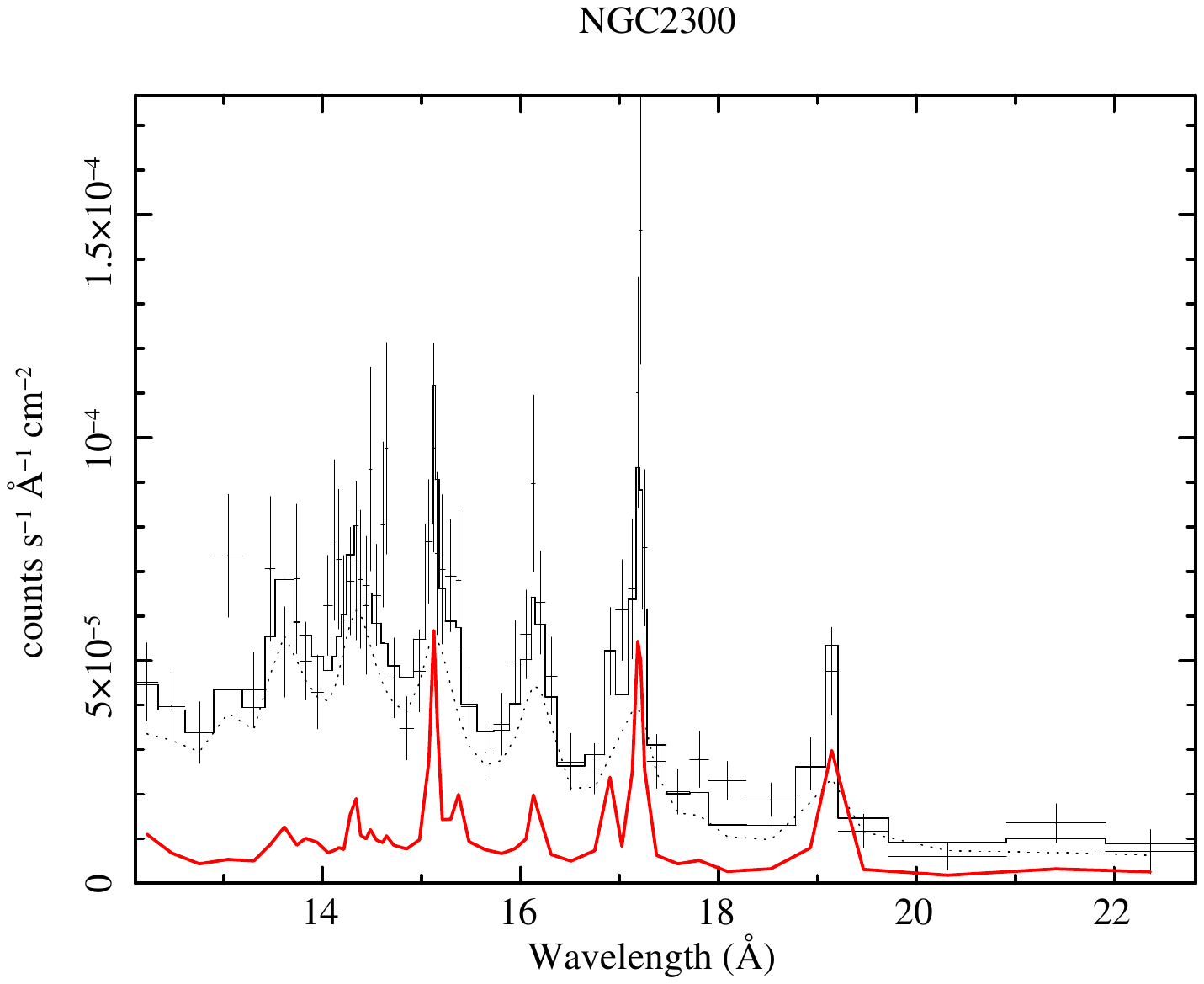} 
\includegraphics[width=0.47\textwidth]{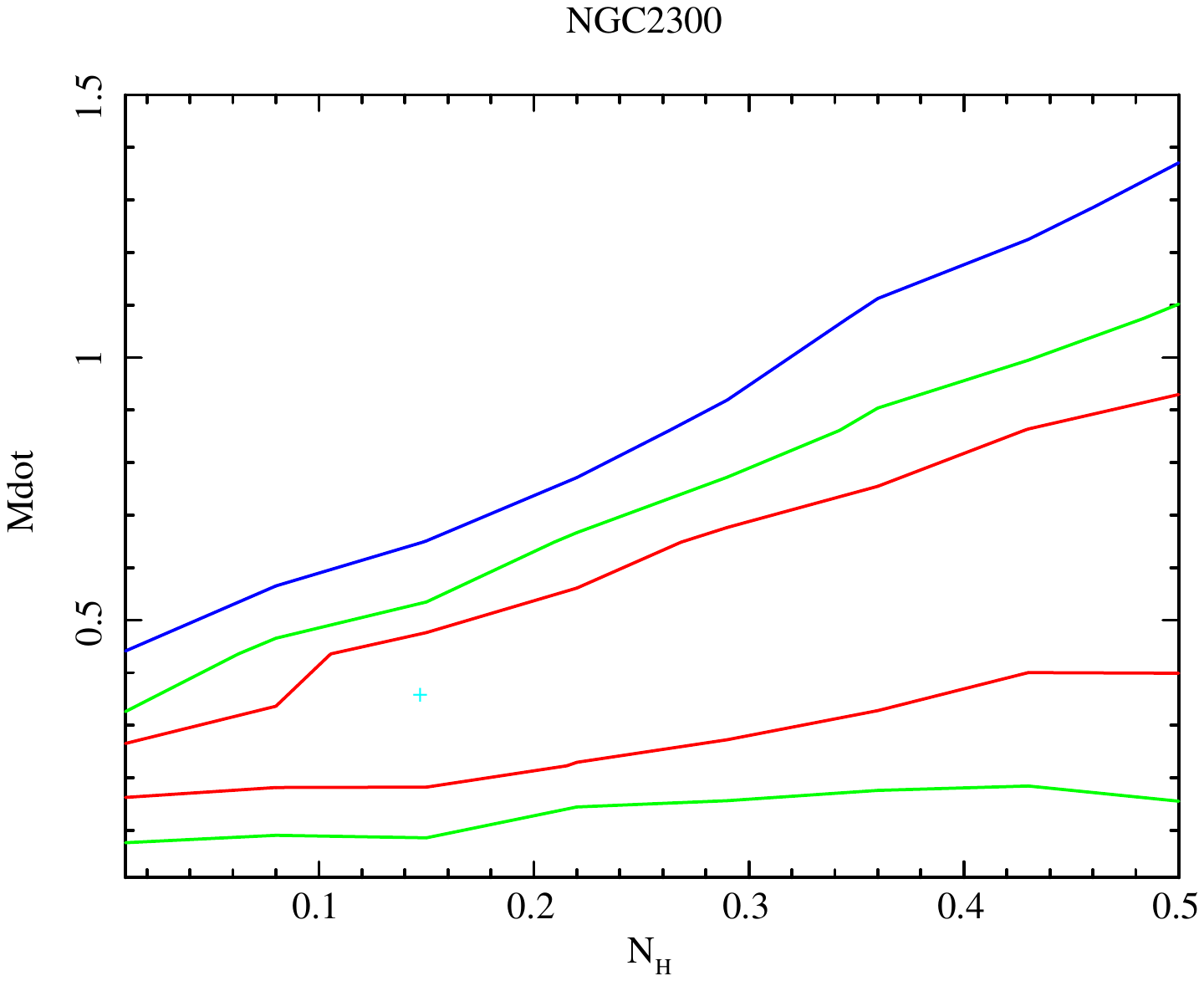}
    \caption{ a) RGS spectrum and b) hidden mass cooling rate versus total interleaved column density.}
\end{figure}

\begin{figure}
\centering  
\includegraphics[width=0.48\textwidth]{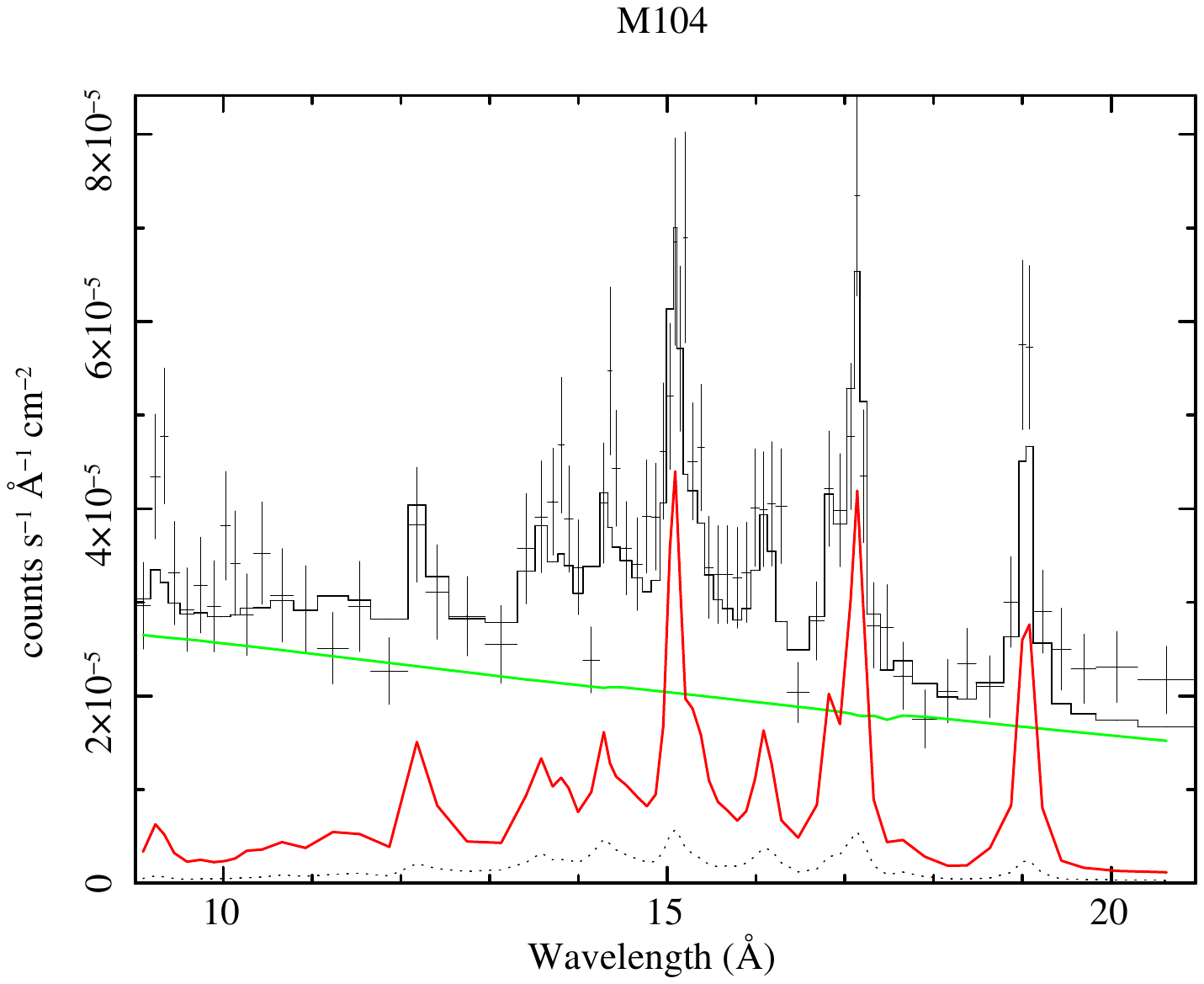}   
\includegraphics[width=0.48\textwidth]{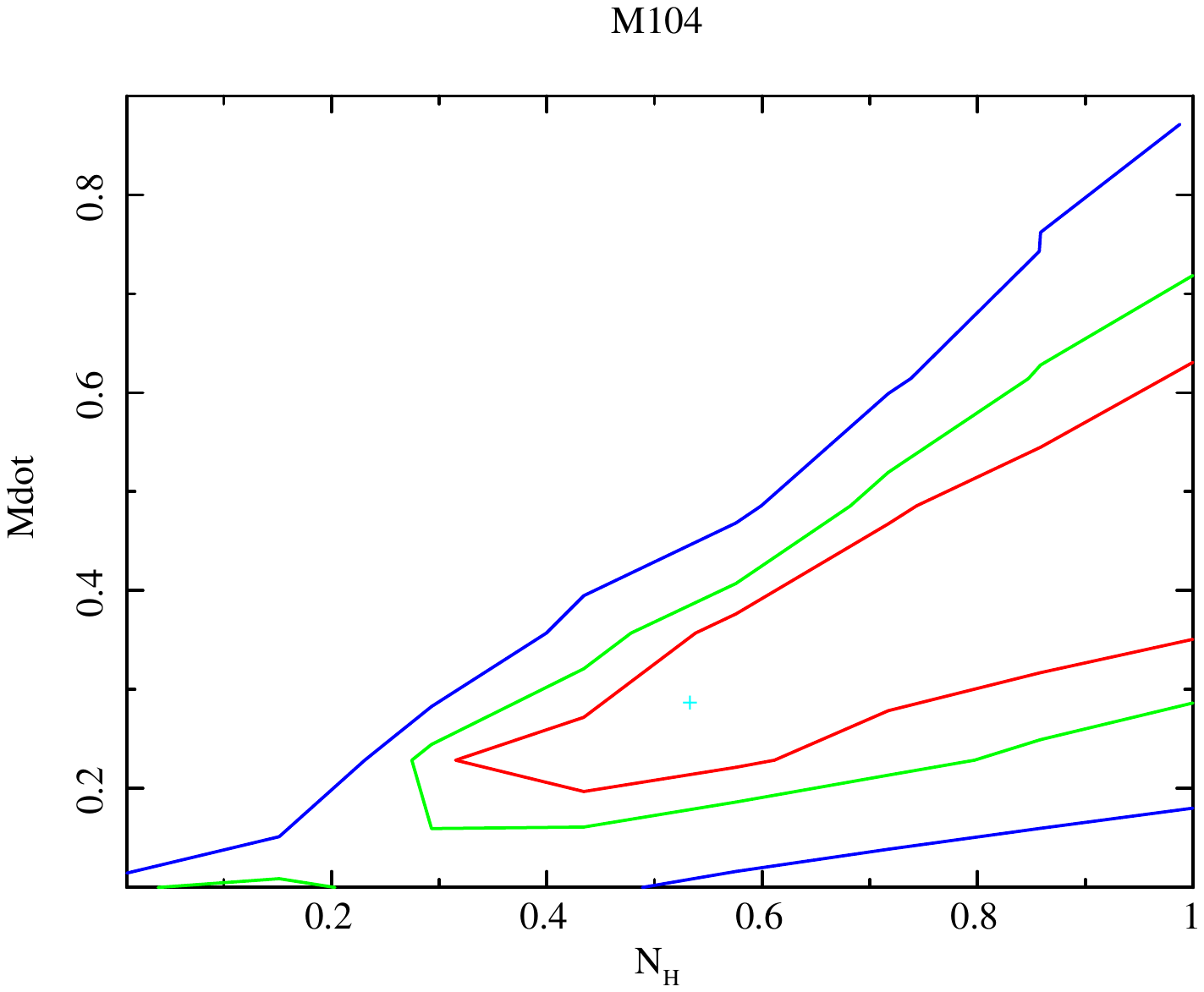} 
    \caption{  a) RGS spectrum and b) hidden mass cooling rate versus total interleaved column density.}
\end{figure}

\begin{figure}
\centering  
\includegraphics[width=0.48\textwidth]{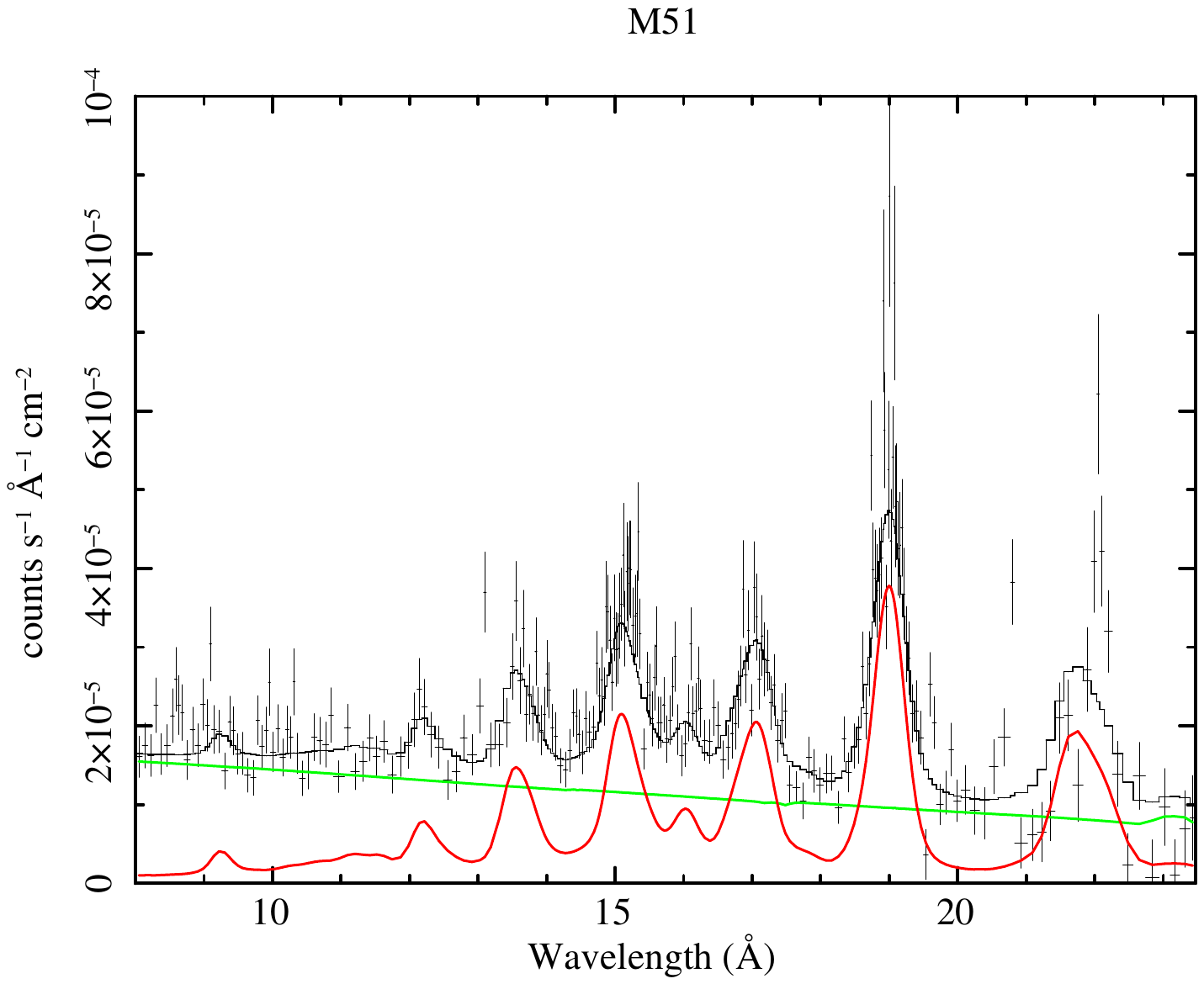}   
\includegraphics[width=0.48\textwidth]{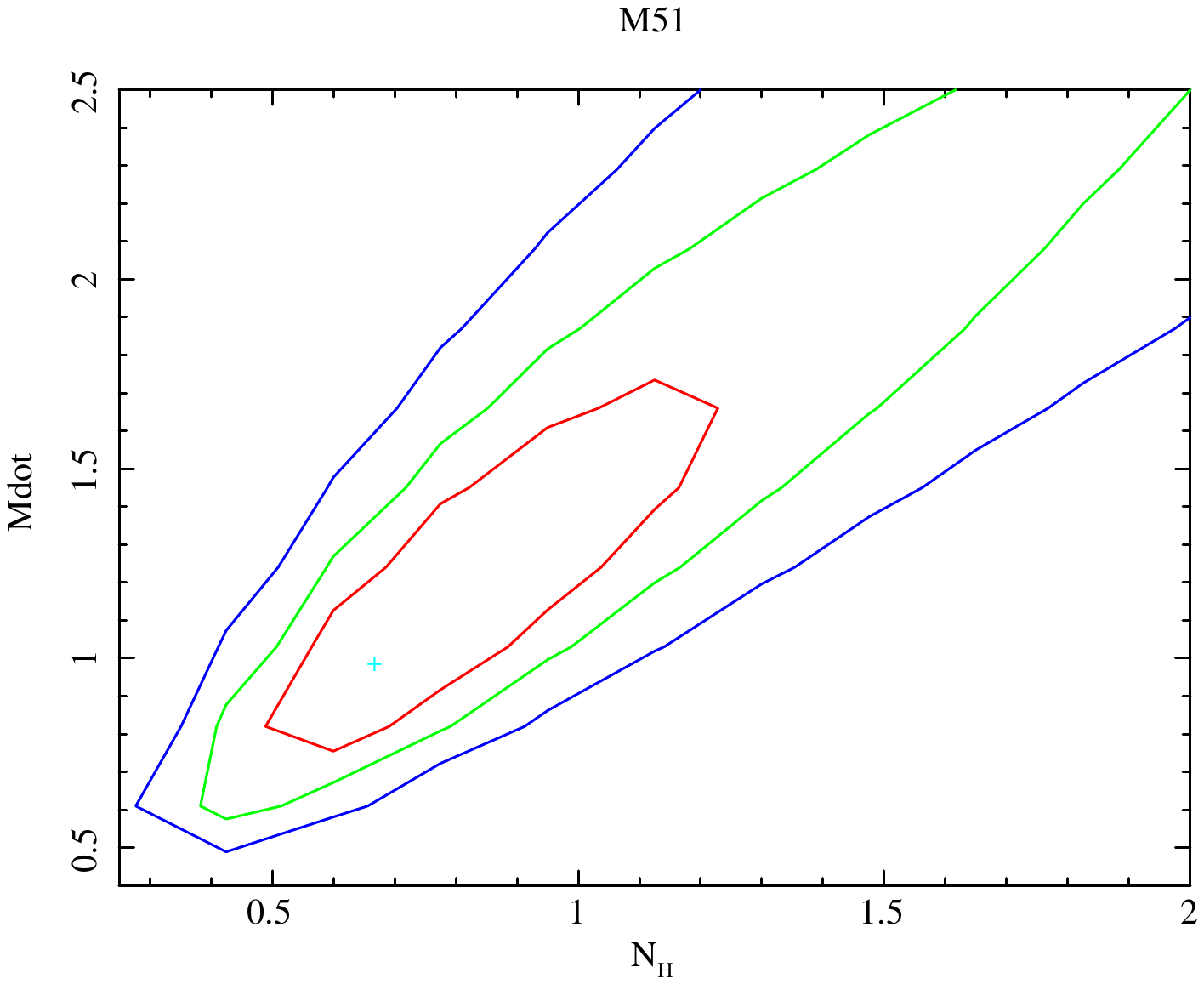} 
    \caption{ a) RGS spectrum and b) hidden mass cooling rate versus total interleaved column density. Note the extra, unfitted, emission line at 24\AA\ which is due to charge exchange \citep{Zhang22}. }
\end{figure}

\begin{figure}
\centering  
\includegraphics[width=0.48\textwidth]{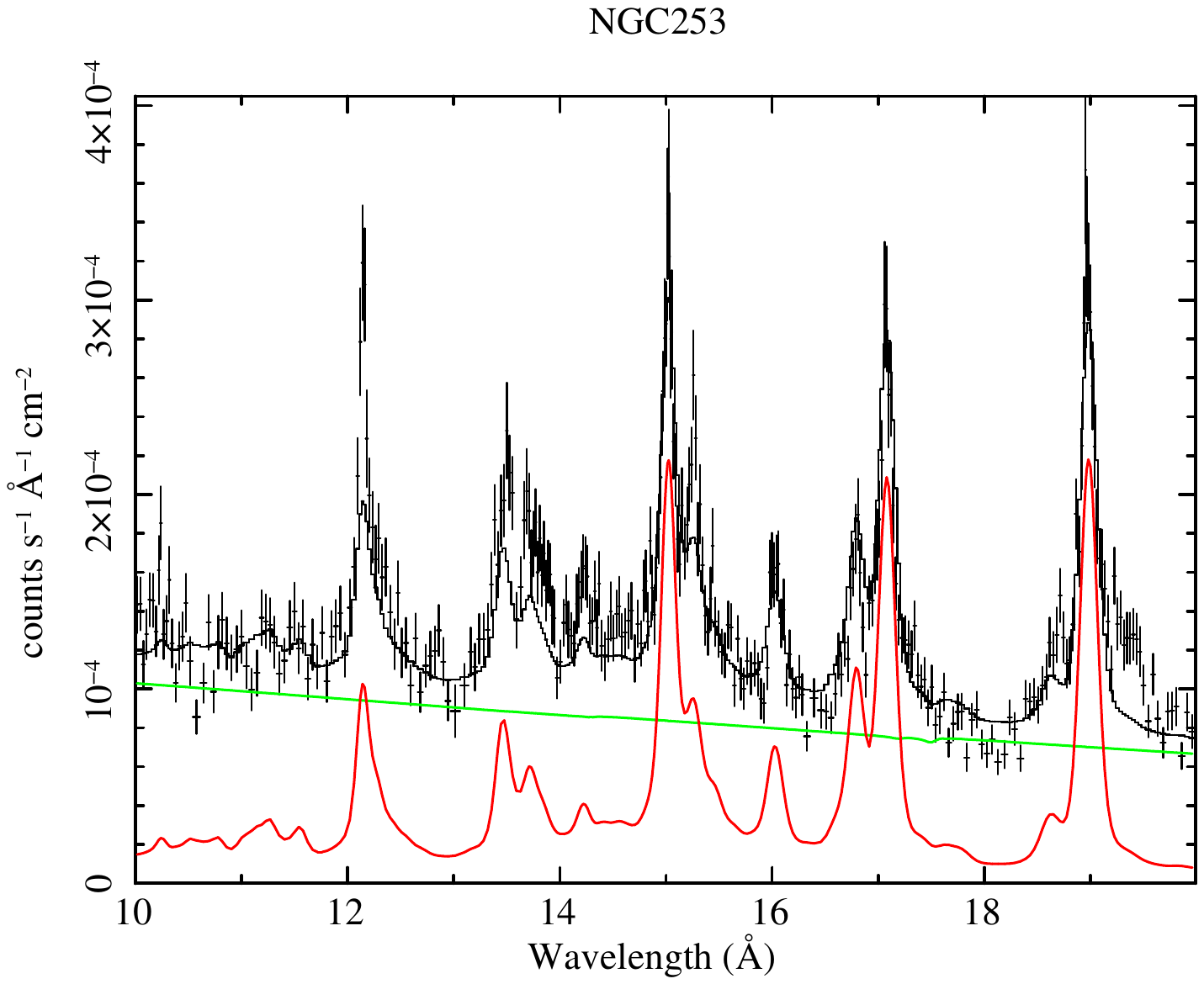}   
\includegraphics[width=0.48\textwidth]{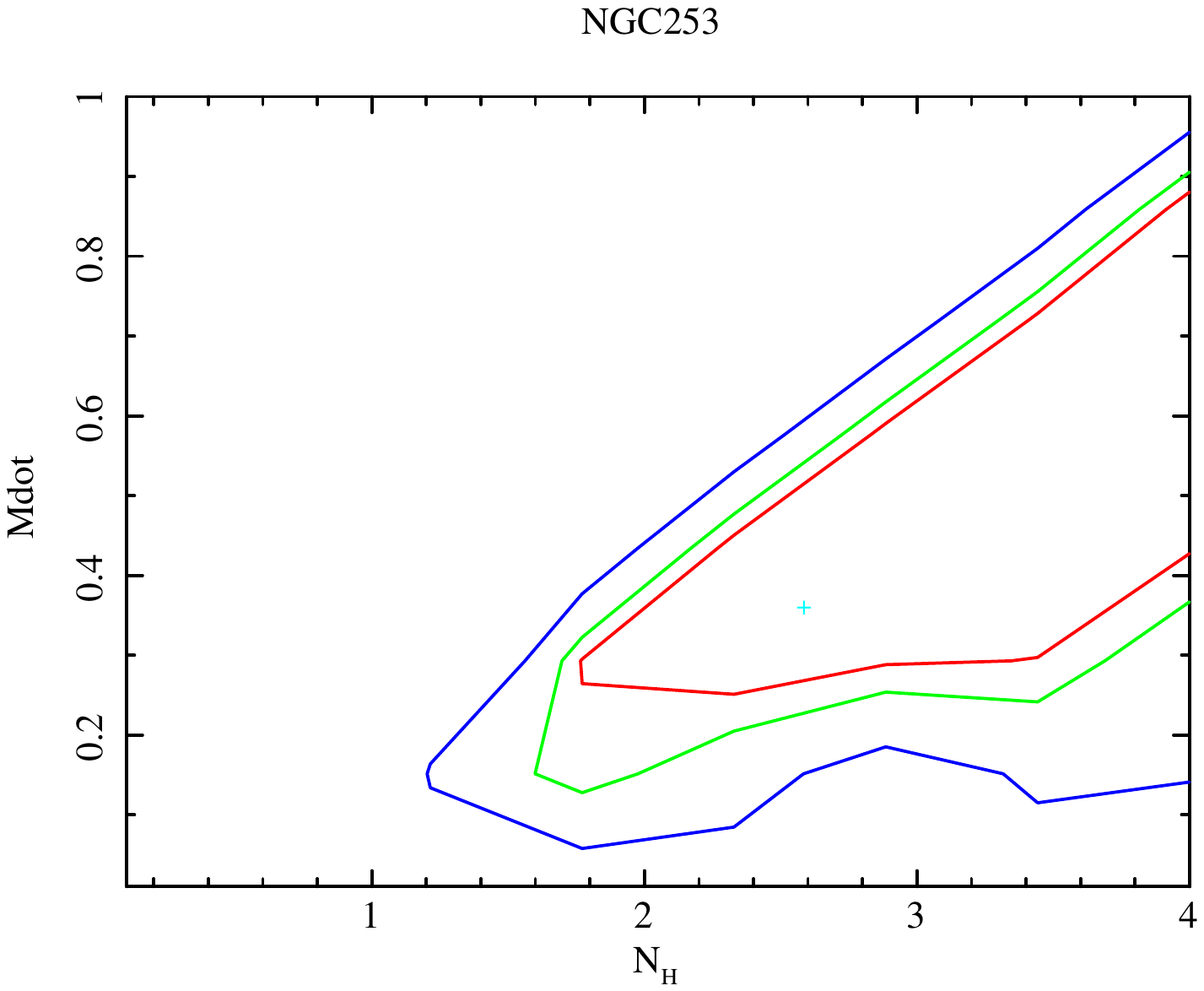} 
    \caption{  a) RGS spectrum and b) hidden mass cooling rate versus total interleaved column density.}
\end{figure}

\section*{Data Availability}

All data presented here are available on the XMM and Chandra archives.



\bibliographystyle{mnras}
\bibliography{cool_core_Mdot_24} 

\begin{thebibliography}{}
\makeatletter
\relax
\def\mn@urlcharsother{\let\do\@makeother \do\$\do\&\do\#\do\^\do\_\do\%\do\~}
\def\mn@doi{\begingroup\mn@urlcharsother \@ifnextchar [ {\mn@doi@} {\mn@doi@[]}}
\def\mn@doi@[#1]#2{\def\@tempa{#1}\ifx\@tempa\@empty \href {http://dx.doi.org/#2} {doi:#2}\else \href {http://dx.doi.org/#2} {#1}\fi \endgroup}
\def\mn@eprint#1#2{\mn@eprint@#1:#2::\@nil}
\def\mn@eprint@arXiv#1{\href {http://arxiv.org/abs/#1} {{\tt arXiv:#1}}}
\def\mn@eprint@dblp#1{\href {http://dblp.uni-trier.de/rec/bibtex/#1.xml} {dblp:#1}}
\def\mn@eprint@#1:#2:#3:#4\@nil{\def\@tempa {#1}\def\@tempb {#2}\def\@tempc {#3}\ifx \@tempc \@empty \let \@tempc \@tempb \let \@tempb \@tempa \fi \ifx \@tempb \@empty \def\@tempb {arXiv}\fi \@ifundefined {mn@eprint@\@tempb}{\@tempb:\@tempc}{\expandafter \expandafter \csname mn@eprint@\@tempb\endcsname \expandafter{\@tempc}}}

\bibitem[\protect\citeauthoryear{{Allen} \& {Fabian}}{{Allen} \& {Fabian}}{1997}]{Allen1997}
{Allen} S.~W.,  {Fabian} A.~C.,  1997, \mn@doi [\mnras] {10.1093/mnras/286.3.583}, \href {https://ui.adsabs.harvard.edu/abs/1997MNRAS.286..583A} {286, 583}

\bibitem[\protect\citeauthoryear{{Arnaud}}{{Arnaud}}{1996}]{Arnaud1996}
{Arnaud} K.~A.,  1996, in {Jacoby} G.~H.,  {Barnes} J.,  eds,  Astronomical Society of the Pacific Conference Series Vol. 101, Astronomical Data Analysis Software and Systems V. p.~17

\bibitem[\protect\citeauthoryear{{Bauer}, {Pietsch}, {Trinchieri}, {Breitschwerdt}, {Ehle}  \& {Read}}{{Bauer} et~al.}{2007}]{Bauer07}
{Bauer} M.,  {Pietsch} W.,  {Trinchieri} G.,  {Breitschwerdt} D.,  {Ehle} M.,   {Read} A.,  2007, \mn@doi [\aap] {10.1051/0004-6361:20066340}, \href {https://ui.adsabs.harvard.edu/abs/2007A&A...467..979B} {467, 979}

\bibitem[\protect\citeauthoryear{{Buote}}{{Buote}}{2017}]{Buote17}
{Buote} D.~A.,  2017, \mn@doi [\apj] {10.3847/1538-4357/834/2/164}, \href {https://ui.adsabs.harvard.edu/abs/2017ApJ...834..164B} {834, 164}

\bibitem[\protect\citeauthoryear{{Cassinelli} \& {Olson}}{{Cassinelli} \& {Olson}}{1979}]{Cassinelli79}
{Cassinelli} J.~P.,  {Olson} G.~L.,  1979, \mn@doi [\apj] {10.1086/156956}, \href {https://ui.adsabs.harvard.edu/abs/1979ApJ...229..304C} {229, 304}

\bibitem[\protect\citeauthoryear{{Cohen}, {Parts}, {Doskoch}, {Wang}, {Petit}, {Leutenegger}  \& {Gagn{\'e}}}{{Cohen} et~al.}{2021}]{Cohen21}
{Cohen} D.~H.,  {Parts} W.,  {Doskoch} G.~M.,  {Wang} J.,  {Petit} V.,  {Leutenegger} M.~A.,   {Gagn{\'e}} M.,  2021, \mn@doi [\mnras] {10.1093/mnras/stab270}, \href {https://ui.adsabs.harvard.edu/abs/2021MNRAS.503..715C} {503, 715}

\bibitem[\protect\citeauthoryear{{Condon}, {Cotton}, {White}, {Legodi}, {Goedhart}, {McAlpine}, {Ratcliffe}  \& {Camilo}}{{Condon} et~al.}{2021}]{Condon21}
{Condon} J.~J.,  {Cotton} W.~D.,  {White} S.~V.,  {Legodi} S.,  {Goedhart} S.,  {McAlpine} K.,  {Ratcliffe} S.~M.,   {Camilo} F.,  2021, \mn@doi [\apj] {10.3847/1538-4357/ac0880}, \href {https://ui.adsabs.harvard.edu/abs/2021ApJ...917...18C} {917, 18}

\bibitem[\protect\citeauthoryear{{Dutta}, {Sharma}  \& {Nelson}}{{Dutta} et~al.}{2022}]{Dutta22}
{Dutta} A.,  {Sharma} P.,   {Nelson} D.,  2022, \mn@doi [\mnras] {10.1093/mnras/stab3653}, \href {https://ui.adsabs.harvard.edu/abs/2022MNRAS.510.3561D} {510, 3561}

\bibitem[\protect\citeauthoryear{{Fabian}}{{Fabian}}{1994}]{Fabian1994rev}
{Fabian} A.~C.,  1994, \mn@doi [\araa] {10.1146/annurev.aa.32.090194.001425}, \href {https://ui.adsabs.harvard.edu/abs/1994ARA&A..32..277F} {32, 277}

\bibitem[\protect\citeauthoryear{{Fabian}}{{Fabian}}{2012}]{Fabian2012}
{Fabian} A.~C.,  2012, \mn@doi [\araa] {10.1146/annurev-astro-081811-125521}, \href {https://ui.adsabs.harvard.edu/abs/2012ARA&A..50..455F} {50, 455}

\bibitem[\protect\citeauthoryear{{Fabian}, {Ferland}, {Sanders}, {McNamara}, {Pinto}  \& {Walker}}{{Fabian} et~al.}{2022}]{Fabian22}
{Fabian} A.~C.,  {Ferland} G.~J.,  {Sanders} J.~S.,  {McNamara} B.~R.,  {Pinto} C.,   {Walker} S.~A.,  2022, \mn@doi [\mnras] {10.1093/mnras/stac2003}, \href {https://ui.adsabs.harvard.edu/abs/2022MNRAS.515.3336F} {515, 3336}

\bibitem[\protect\citeauthoryear{{Fabian}, {Sanders}, {Ferland}, {McNamara}, {Pinto}  \& {Walker}}{{Fabian} et~al.}{2023a}]{Fabian23}
{Fabian} A.~C.,  {Sanders} J.~S.,  {Ferland} G.~J.,  {McNamara} B.~R.,  {Pinto} C.,   {Walker} S.~A.,  2023a, \mn@doi [\mnras] {10.1093/mnras/stad507}, \href {https://ui.adsabs.harvard.edu/abs/2023MNRAS.521.1794F} {521, 1794}

\bibitem[\protect\citeauthoryear{{Fabian}, {Sanders}, {Ferland}, {McNamara}, {Pinto}  \& {Walker}}{{Fabian} et~al.}{2023b}]{Fabian23b}
{Fabian} A.~C.,  {Sanders} J.~S.,  {Ferland} G.~J.,  {McNamara} B.~R.,  {Pinto} C.,   {Walker} S.~A.,  2023b, \mn@doi [\mnras] {10.1093/mnras/stad1870}, \href {https://ui.adsabs.harvard.edu/abs/2023MNRAS.524..716F} {524, 716}

\bibitem[\protect\citeauthoryear{{Fabian}, {Sanders}, {Ferland}, {McNamara}, {Pinto}  \& {Walker}}{{Fabian} et~al.}{2024a}]{Fabian24}
{Fabian} A.~C.,  {Sanders} J.~S.,  {Ferland} G.~J.,  {McNamara} B.~R.,  {Pinto} C.,   {Walker} S.~A.,  2024a, \mn@doi [\mnras] {10.1093/mnras/stae1206}, \href {https://ui.adsabs.harvard.edu/abs/2024MNRAS.531..267F} {531, 267}

\bibitem[\protect\citeauthoryear{{Fabian} et~al.,}{{Fabian} et~al.}{2024b}]{Fabian24b}
{Fabian} A.~C.,  et~al., 2024b, \mn@doi [\mnras] {10.1093/mnras/stae2414}, \href {https://ui.adsabs.harvard.edu/abs/2024MNRAS.535.2173F} {535, 2173}

\bibitem[\protect\citeauthoryear{{Gliozzi}, {Sambruna}  \& {Brandt}}{{Gliozzi} et~al.}{2003}]{Gliozzi03A&A}
{Gliozzi} M.,  {Sambruna} R.~M.,   {Brandt} W.~N.,  2003, \mn@doi [\aap] {10.1051/0004-6361:20031050}, \href {https://ui.adsabs.harvard.edu/abs/2003A&A...408..949G} {408, 949}

\bibitem[\protect\citeauthoryear{{Grossov{\'a}} et~al.,}{{Grossov{\'a}} et~al.}{2019}]{Grossova19}
{Grossov{\'a}} R.,  et~al., 2019, \mn@doi [\mnras] {10.1093/mnras/stz1728}, \href {https://ui.adsabs.harvard.edu/abs/2019MNRAS.488.1917G} {488, 1917}

\bibitem[\protect\citeauthoryear{{Grossov{\'a}} et~al.,}{{Grossov{\'a}} et~al.}{2022}]{Grossova2022}
{Grossov{\'a}} R.,  et~al., 2022, \mn@doi [\apjs] {10.3847/1538-4365/ac366c}, \href {https://ui.adsabs.harvard.edu/abs/2022ApJS..258...30G} {258, 30}

\bibitem[\protect\citeauthoryear{{Gu}, {Greene}, {Newman}, {Kreisch}, {Quenneville}, {Ma}  \& {Blakeslee}}{{Gu} et~al.}{2022}]{Gu22}
{Gu} M.,  {Greene} J.~E.,  {Newman} A.~B.,  {Kreisch} C.,  {Quenneville} M.~E.,  {Ma} C.-P.,   {Blakeslee} J.~P.,  2022, \mn@doi [\apj] {10.3847/1538-4357/ac69ea}, \href {https://ui.adsabs.harvard.edu/abs/2022ApJ...932..103G} {932, 103}

\bibitem[\protect\citeauthoryear{{Ivey}, {Fabian}, {Sanders}, {Pinto}, {Ferland}, {Walker}  \& {Jiang}}{{Ivey} et~al.}{2024}]{Ivey24}
{Ivey} L.~R.,  {Fabian} A.~C.,  {Sanders} J.~S.,  {Pinto} C.,  {Ferland} G.~J.,  {Walker} S.,   {Jiang} J.,  2024, \mn@doi [\mnras] {10.1093/mnras/stae2516}, \href {https://ui.adsabs.harvard.edu/abs/2024MNRAS.535.2697I} {535, 2697}

\bibitem[\protect\citeauthoryear{{Kaastra}, {Ferrigno}, {Tamura}, {Paerels}, {Peterson}  \& {Mittaz}}{{Kaastra} et~al.}{2001}]{Kaastra2001}
{Kaastra} J.~S.,  {Ferrigno} C.,  {Tamura} T.,  {Paerels} F.~B.~S.,  {Peterson} J.~R.,   {Mittaz} J.~P.~D.,  2001, \mn@doi [\aap] {10.1051/0004-6361:20000041}, \href {https://ui.adsabs.harvard.edu/abs/2001A&A...365L..99K} {365, L99}

\bibitem[\protect\citeauthoryear{{Khosroshahi}, {Jones}  \& {Ponman}}{{Khosroshahi} et~al.}{2004}]{Khosroshahi04}
{Khosroshahi} H.~G.,  {Jones} L.~R.,   {Ponman} T.~J.,  2004, \mn@doi [\mnras] {10.1111/j.1365-2966.2004.07575.x}, \href {https://ui.adsabs.harvard.edu/abs/2004MNRAS.349.1240K} {349, 1240}

\bibitem[\protect\citeauthoryear{{Kim} et~al.,}{{Kim} et~al.}{2020}]{Kim20}
{Kim} D.-W.,  et~al., 2020, \mn@doi [\mnras] {10.1093/mnras/stz3530}, \href {https://ui.adsabs.harvard.edu/abs/2020MNRAS.492.2095K} {492, 2095}

\bibitem[\protect\citeauthoryear{{Krumholz}, {Myers}, {Klein}  \& {McKee}}{{Krumholz} et~al.}{2016}]{Krumholz16}
{Krumholz} M.~R.,  {Myers} A.~T.,  {Klein} R.~I.,   {McKee} C.~F.,  2016, \mn@doi [\mnras] {10.1093/mnras/stw1236}, \href {https://ui.adsabs.harvard.edu/abs/2016MNRAS.460.3272K} {460, 3272}

\bibitem[\protect\citeauthoryear{{Lakhchaura} et~al.,}{{Lakhchaura} et~al.}{2018}]{Lakhchaura2018}
{Lakhchaura} K.,  et~al., 2018, \mn@doi [\mnras] {10.1093/mnras/sty2565}, \href {https://ui.adsabs.harvard.edu/abs/2018MNRAS.481.4472L} {481, 4472}

\bibitem[\protect\citeauthoryear{{Leutenegger}, {Cohen}, {Zsarg{\'o}}, {Martell}, {MacArthur}, {Owocki}, {Gagn{\'e}}  \& {Hillier}}{{Leutenegger} et~al.}{2010}]{Leutenegger10}
{Leutenegger} M.~A.,  {Cohen} D.~H.,  {Zsarg{\'o}} J.,  {Martell} E.~M.,  {MacArthur} J.~P.,  {Owocki} S.~P.,  {Gagn{\'e}} M.,   {Hillier} D.~J.,  2010, \mn@doi [\apj] {10.1088/0004-637X/719/2/1767}, \href {https://ui.adsabs.harvard.edu/abs/2010ApJ...719.1767L} {719, 1767}

\bibitem[\protect\citeauthoryear{{Li}, {Wang}  \& {Hameed}}{{Li} et~al.}{2007}]{Li07}
{Li} Z.,  {Wang} Q.~D.,   {Hameed} S.,  2007, \mn@doi [\mnras] {10.1111/j.1365-2966.2007.11513.x}, \href {https://ui.adsabs.harvard.edu/abs/2007MNRAS.376..960L} {376, 960}

\bibitem[\protect\citeauthoryear{{Li} et~al.,}{{Li} et~al.}{2011}]{Li11}
{Li} Z.,  et~al., 2011, \mn@doi [\apj] {10.1088/0004-637X/730/2/84}, \href {https://ui.adsabs.harvard.edu/abs/2011ApJ...730...84L} {730, 84}

\bibitem[\protect\citeauthoryear{{Liu}, {Pinto}, {Fabian}, {Russell}  \& {Sanders}}{{Liu} et~al.}{2019}]{Liu2019}
{Liu} H.,  {Pinto} C.,  {Fabian} A.~C.,  {Russell} H.~R.,   {Sanders} J.~S.,  2019, \mn@doi [\mnras] {10.1093/mnras/stz456}, \href {https://ui.adsabs.harvard.edu/abs/2019MNRAS.485.1757L} {485, 1757}

\bibitem[\protect\citeauthoryear{{Lopez}, {Lopez}, {Nguyen}, {Thompson}, {Mathur}, {Bolatto}, {Vulic}  \& {Sardone}}{{Lopez} et~al.}{2023}]{Lopez23}
{Lopez} S.,  {Lopez} L.~A.,  {Nguyen} D.~D.,  {Thompson} T.~A.,  {Mathur} S.,  {Bolatto} A.~D.,  {Vulic} N.,   {Sardone} A.,  2023, \mn@doi [\apj] {10.3847/1538-4357/aca65e}, \href {https://ui.adsabs.harvard.edu/abs/2023ApJ...942..108L} {942, 108}

\bibitem[\protect\citeauthoryear{{McDonald}, {Veilleux}  \& {Rupke}}{{McDonald} et~al.}{2012}]{McD12}
{McDonald} M.,  {Veilleux} S.,   {Rupke} D. S.~N.,  2012, \mn@doi [\apj] {10.1088/0004-637X/746/2/153}, \href {https://ui.adsabs.harvard.edu/abs/2012ApJ...746..153M} {746, 153}

\bibitem[\protect\citeauthoryear{{McDonald}, {Gaspari}, {McNamara}  \& {Tremblay}}{{McDonald} et~al.}{2018}]{McDonald2018}
{McDonald} M.,  {Gaspari} M.,  {McNamara} B.~R.,   {Tremblay} G.~R.,  2018, \mn@doi [\apj] {10.3847/1538-4357/aabace}, \href {https://ui.adsabs.harvard.edu/abs/2018ApJ...858...45M} {858, 45}

\bibitem[\protect\citeauthoryear{{McNamara} \& {Nulsen}}{{McNamara} \& {Nulsen}}{2012}]{McNamara2012}
{McNamara} B.~R.,  {Nulsen} P.~E.~J.,  2012, \mn@doi [New Journal of Physics] {10.1088/1367-2630/14/5/055023}, \href {https://ui.adsabs.harvard.edu/abs/2012NJPh...14e5023M} {14, 055023}

\bibitem[\protect\citeauthoryear{{Navarro}, {Frenk}  \& {White}}{{Navarro} et~al.}{1995}]{Navarro95}
{Navarro} J.~F.,  {Frenk} C.~S.,   {White} S. D.~M.,  1995, \mn@doi [\mnras] {10.1093/mnras/275.3.720}, \href {https://ui.adsabs.harvard.edu/abs/1995MNRAS.275..720N} {275, 720}

\bibitem[\protect\citeauthoryear{{O'Sullivan}, {Vrtilek}, {Harris}  \& {Ponman}}{{O'Sullivan} et~al.}{2007}]{Osullivan07}
{O'Sullivan} E.,  {Vrtilek} J.~M.,  {Harris} D.~E.,   {Ponman} T.~J.,  2007, \mn@doi [\apj] {10.1086/511778}, \href {https://ui.adsabs.harvard.edu/abs/2007ApJ...658..299O} {658, 299}

\bibitem[\protect\citeauthoryear{{Oldham} \& {Auger}}{{Oldham} \& {Auger}}{2018}]{Oldham2018}
{Oldham} L.,  {Auger} M.,  2018, \mn@doi [\mnras] {10.1093/mnras/stx2969}, \href {https://ui.adsabs.harvard.edu/abs/2018MNRAS.474.4169O} {474, 4169}

\bibitem[\protect\citeauthoryear{{Owen} \& {Warwick}}{{Owen} \& {Warwick}}{2009}]{Owen09}
{Owen} R.~A.,  {Warwick} R.~S.,  2009, \mn@doi [\mnras] {10.1111/j.1365-2966.2009.14464.x}, \href {https://ui.adsabs.harvard.edu/abs/2009MNRAS.394.1741O} {394, 1741}

\bibitem[\protect\citeauthoryear{{Panagoulia}, {Fabian}, {Sanders}  \& {Hlavacek-Larrondo}}{{Panagoulia} et~al.}{2014}]{Panagoulia2014}
{Panagoulia} E.~K.,  {Fabian} A.~C.,  {Sanders} J.~S.,   {Hlavacek-Larrondo} J.,  2014, \mn@doi [\mnras] {10.1093/mnras/stu1499}, \href {https://ui.adsabs.harvard.edu/abs/2014MNRAS.444.1236P} {444, 1236}

\bibitem[\protect\citeauthoryear{{Pellegrini}, {Venturi}, {Comastri}, {Fabbiano}, {Fiore}, {Vignali}, {Morganti}  \& {Trinchieri}}{{Pellegrini} et~al.}{2003a}]{Pellegrini03}
{Pellegrini} S.,  {Venturi} T.,  {Comastri} A.,  {Fabbiano} G.,  {Fiore} F.,  {Vignali} C.,  {Morganti} R.,   {Trinchieri} G.,  2003a, \mn@doi [\apj] {10.1086/346184}, \href {https://ui.adsabs.harvard.edu/abs/2003ApJ...585..677P} {585, 677}

\bibitem[\protect\citeauthoryear{{Pellegrini}, {Baldi}, {Fabbiano}  \& {Kim}}{{Pellegrini} et~al.}{2003b}]{Pellegrini03a}
{Pellegrini} S.,  {Baldi} A.,  {Fabbiano} G.,   {Kim} D.~W.,  2003b, \mn@doi [\apj] {10.1086/378235}, \href {https://ui.adsabs.harvard.edu/abs/2003ApJ...597..175P} {597, 175}

\bibitem[\protect\citeauthoryear{{Peterson} \& {Fabian}}{{Peterson} \& {Fabian}}{2006}]{Peterson2006}
{Peterson} J.~R.,  {Fabian} A.~C.,  2006, \mn@doi [\physrep] {10.1016/j.physrep.2005.12.007}, \href {https://ui.adsabs.harvard.edu/abs/2006PhR...427....1P} {427, 1}

\bibitem[\protect\citeauthoryear{{Peterson} et~al.,}{{Peterson} et~al.}{2001}]{Peterson2001}
{Peterson} J.~R.,  et~al., 2001, \mn@doi [\aap] {10.1051/0004-6361:20000021}, \href {https://ui.adsabs.harvard.edu/abs/2001A&A...365L.104P} {365, L104}

\bibitem[\protect\citeauthoryear{{Peterson}, {Kahn}, {Paerels}, {Kaastra}, {Tamura}, {Bleeker}, {Ferrigno}  \& {Jernigan}}{{Peterson} et~al.}{2003}]{Peterson2003}
{Peterson} J.~R.,  {Kahn} S.~M.,  {Paerels} F.~B.~S.,  {Kaastra} J.~S.,  {Tamura} T.,  {Bleeker} J.~A.~M.,  {Ferrigno} C.,   {Jernigan} J.~G.,  2003, \mn@doi [\apj] {10.1086/374830}, \href {https://ui.adsabs.harvard.edu/abs/2003ApJ...590..207P} {590, 207}

\bibitem[\protect\citeauthoryear{{Pinto} et~al.,}{{Pinto} et~al.}{2016}]{Pinto2016}
{Pinto} C.,  et~al., 2016, \mn@doi [\mnras] {10.1093/mnras/stw1444}, \href {https://ui.adsabs.harvard.edu/abs/2016MNRAS.461.2077P} {461, 2077}

\bibitem[\protect\citeauthoryear{{Russell}, {Ponman}  \& {Sanderson}}{{Russell} et~al.}{2007}]{Russell07}
{Russell} P.~A.,  {Ponman} T.~J.,   {Sanderson} A. J.~R.,  2007, \mn@doi [\mnras] {10.1111/j.1365-2966.2007.11660.x}, \href {https://ui.adsabs.harvard.edu/abs/2007MNRAS.378.1217R} {378, 1217}

\bibitem[\protect\citeauthoryear{{Stern}, {Fielding}, {Faucher-Gigu{\`e}re}  \& {Quataert}}{{Stern} et~al.}{2019}]{Stern19}
{Stern} J.,  {Fielding} D.,  {Faucher-Gigu{\`e}re} C.-A.,   {Quataert} E.,  2019, \mn@doi [\mnras] {10.1093/mnras/stz1859}, \href {https://ui.adsabs.harvard.edu/abs/2019MNRAS.488.2549S} {488, 2549}

\bibitem[\protect\citeauthoryear{{Stewart} \& {Fabian}}{{Stewart} \& {Fabian}}{1981}]{Stewart81}
{Stewart} G.~C.,  {Fabian} A.~C.,  1981, \mn@doi [\mnras] {10.1093/mnras/197.3.713}, \href {https://ui.adsabs.harvard.edu/abs/1981MNRAS.197..713S} {197, 713}

\bibitem[\protect\citeauthoryear{{Sultan}, {Faucher-Gigu{\`e}re}, {Stern}, {Rotshtein}, {Byrne}  \& {Wijers}}{{Sultan} et~al.}{2025}]{Sultan25}
{Sultan} I.,  {Faucher-Gigu{\`e}re} C.-A.,  {Stern} J.,  {Rotshtein} S.,  {Byrne} L.,   {Wijers} N.,  2025, \mn@doi [\mnras] {10.1093/mnras/staf786}, \href {https://ui.adsabs.harvard.edu/abs/2025MNRAS.540.1017S} {540, 1017}

\bibitem[\protect\citeauthoryear{{Tamhane} et~al.,}{{Tamhane} et~al.}{2025}]{Tamhane25}
{Tamhane} P.,  et~al., 2025, \mn@doi [arXiv e-prints] {10.48550/arXiv.2507.13431}, \href {https://ui.adsabs.harvard.edu/abs/2025arXiv250713431T} {p. arXiv:2507.13431}

\bibitem[\protect\citeauthoryear{{Tashiro}, {Maejima}, {Toda}, {Kelley}, {Reichenthal}, {Lobell}, {Petre}  \& {et al.}}{{Tashiro} et~al.}{2018}]{XRISM2018}
{Tashiro} M.,  {Maejima} H.,  {Toda} K.,  {Kelley} R.,  {Reichenthal} L.,  {Lobell} J.,  {Petre} R.,   {et al.} 2018, in {den Herder} J.-W.~A.,  {Nikzad} S.,   {Nakazawa} K.,  eds,  Society of Photo-Optical Instrumentation Engineers (SPIE) Conference Series Vol. 10699, Space Telescopes and Instrumentation 2018: Ultraviolet to Gamma Ray. p. 1069922, \mn@doi{10.1117/12.2309455}

\bibitem[\protect\citeauthoryear{{Tumlinson}, {Peeples}  \& {Werk}}{{Tumlinson} et~al.}{2017}]{Tumlinson17ARA&A..55..389T}
{Tumlinson} J.,  {Peeples} M.~S.,   {Werk} J.~K.,  2017, \mn@doi [\araa] {10.1146/annurev-astro-091916-055240}, \href {https://ui.adsabs.harvard.edu/abs/2017ARA&A..55..389T} {55, 389}

\bibitem[\protect\citeauthoryear{{Worrall}, {Birkinshaw}, {O'Sullivan}, {Zezas}, {Wolter}, {Trinchieri}  \& {Fabbiano}}{{Worrall} et~al.}{2010}]{Worrall10}
{Worrall} D.~M.,  {Birkinshaw} M.,  {O'Sullivan} E.,  {Zezas} A.,  {Wolter} A.,  {Trinchieri} G.,   {Fabbiano} G.,  2010, \mn@doi [\mnras] {10.1111/j.1365-2966.2010.17162.x}, \href {https://ui.adsabs.harvard.edu/abs/2010MNRAS.408..701W} {408, 701}

\bibitem[\protect\citeauthoryear{{Zezas}, {Birkinshaw}, {Worrall}, {Peters}  \& {Fabbiano}}{{Zezas} et~al.}{2005}]{Zezas05}
{Zezas} A.,  {Birkinshaw} M.,  {Worrall} D.~M.,  {Peters} A.,   {Fabbiano} G.,  2005, \mn@doi [\apj] {10.1086/430044}, \href {https://ui.adsabs.harvard.edu/abs/2005ApJ...627..711Z} {627, 711}

\bibitem[\protect\citeauthoryear{{Zhang}, {Wang}, {Sun}, {Long}, {Sun}  \& {Ji}}{{Zhang} et~al.}{2022}]{Zhang22}
{Zhang} S.,  {Wang} Q.~D.,  {Sun} W.,  {Long} M.,  {Sun} J.,   {Ji} L.,  2022, \mn@doi [\apj] {10.3847/1538-4357/aca01a}, \href {https://ui.adsabs.harvard.edu/abs/2022ApJ...941...68Z} {941, 68}

\bibitem[\protect\citeauthoryear{{den Herder} et~al.,}{{den Herder} et~al.}{2001}]{denHerder2001}
{den Herder} J.~W.,  et~al., 2001, \mn@doi [\aap] {10.1051/0004-6361:20000058}, \href {https://ui.adsabs.harvard.edu/abs/2001A&A...365L...7D} {365, L7}

\bibitem[\protect\citeauthoryear{{van Dokkum} \& {Conroy}}{{van Dokkum} \& {Conroy}}{2010}]{vDC2010}
{van Dokkum} P.~G.,  {Conroy} C.,  2010, \mn@doi [\nat] {10.1038/nature09578}, \href {https://ui.adsabs.harvard.edu/abs/2010Natur.468..940V} {468, 940}

\bibitem[\protect\citeauthoryear{{van Dokkum} \& {Conroy}}{{van Dokkum} \& {Conroy}}{2024}]{VanDokkum24}
{van Dokkum} P.,  {Conroy} C.,  2024, \mn@doi [arXiv e-prints] {10.48550/arXiv.2407.06281}, \href {https://ui.adsabs.harvard.edu/abs/2024arXiv240706281V} {p. arXiv:2407.06281}

\makeatother
\end{thebibliography}





\bsp	
\label{lastpage}
\end{document}